\documentclass[journal]{aa}

\usepackage{graphicx}
\usepackage{amsmath}
\usepackage{amssymb}
\usepackage{verbatim}
\usepackage{array}
\usepackage{txfonts}
\usepackage{natbib}
\usepackage[switch,pagewise]{lineno}
\usepackage{multirow}

\bibpunct{(}{)}{;}{a}{}{,}

\newcommand{\celltspace}{\rule{0pt}{2.8ex}}
\newcommand{\cellbspace}{\rule[-1.4ex]{0pt}{0pt}}

\def\deg{\ensuremath{^\circ}}

\defcitealias{Chan:1993}{CL93}
\defcitealias{Higdon:2009}{HLR09}
\defcitealias{Jean:2009}{JGMF09}

\begin{document}

\title{Galactic annihilation emission from nucleosynthesis positrons}
\titlerunning{Galactic annihilation emission from nucleosynthesis positrons}
\author{P.~Martin$^{1,2}$ \and A.~W.~Strong$^{1}$ \and P.~Jean$^{3}$ \and A.~Alexis$^{3}$ \and R.~Diehl$^{1}$}
\authorrunning{P.~Martin et~al.}
\institute{Max Planck Institut f\"ur extraterrestrische Physik (MPE), Postfach 1312, 85741 Garching, Germany 
	      \and UJF/CNRS, Institut de Plan\'etologie et d'Astrophysique de Grenoble (IPAG), UMR5274, BP53, 38041 Grenoble cedex 9, France
              \and UPS/CNRS, Institut de Recherche en Astrophysique et Plan\'etologie (IRAP), UMR5277, BP44346, 31028 Toulouse cedex 4, France}

\date{Received 22 December 2011 / Accepted 25 April 2012}
\abstract{The Galaxy hosts a widespread population of low-energy positrons revealed by successive generations of gamma-ray telescopes through a bright annihilation emission from the bulge region, with a fainter contribution from the inner disk. The exact origin of these particles remains currently unknown.}{We estimate the contribution to the annihilation signal of positrons generated in the decay of radioactive $^{26}$Al, $^{56}$Ni and $^{44}$Ti.}{We adapted the GALPROP propagation code to simulate the transport and annihilation of radioactivity positrons in a model of our Galaxy. Using plausible source spatial distributions, we explored several possible propagation scenarios to account for the large uncertainties on the transport of $\sim$\,MeV positrons in the interstellar medium. We then compared the predicted intensity distributions to the INTEGRAL/SPI observations.}{We obtain similar intensity distributions with small bulge-to-disk ratios, even for extreme large-scale transport prescriptions. At least half of the positrons annihilate close to their sources, even when they are allowed to travel far away. In the high-diffusion, ballistic case, up to 40\% of them escape the Galaxy. In proportion, this affects bulge positrons more than disk positrons because they are injected further off the plane in a tenuous medium, while disk positrons are mostly injected in the dense molecular ring. The predicted intensity distributions are fully consistent with the observed longitudinally-extended disk-like emission, but the transport scenario cannot be strongly constrained by the current data.}{Nucleosynthesis positrons alone cannot account for the observed annihilation emission in the frame of our model. An additional component is needed to explain the strong bulge contribution, and the latter is very likely concentrated in the central regions if positrons have initial energies in the 100\,keV-1\,MeV range.}
\maketitle

\section{Introduction}
\label{intro}


\indent Over the past decades, it has become clear that our Galaxy can produce substantial amounts of antimatter and give rise to several large-scale populations of antiparticles that coexist with our matter environment at the current cosmic time. In particular, a sizeable number of positrons apparently fill the entire Galaxy from the very centre to the peripheral regions.\\
\indent The Galactic population of positrons is usually divided into low-energy (typically $\leq$10\,MeV) and high-energy (typically $\geq$100\,MeV) particles, and it seems that there is more than historical or experimental reasons to do that \citep{Prantzos:2011}. While high-energy positrons can be observed directly as a cosmic-ray component in the interplanetary medium, low-energy positrons are revealed indirectly through gamma-ray observations of the sky in the $\sim$100\,keV-1\,MeV range. The latter have indeed shown an unambiguous signature of electron-positron annihilation from the inner Galactic regions, in the form of a line at 511\,keV and a continuum below.\\
\indent That our Galaxy can produce and host a substantial population of positrons is actually not a surprise. There are many physical processes able to provide non-thermal positrons over a broad range of energies (photon-photon pair production, radioactive decay, hadronic interactions, ...), and as many astrophysical phenomena likely to be the sites of one or more of these processes (pulsars, X-ray binaries, supernovae, cosmic rays, ...). If we add more speculative channels like dark matter particles creating positrons through their annihilation or decay, it becomes clear that the real challenge of positron astrophysics is not to find \textit{an} origin for these particles, but to identify \textit{the} source or combination or sources that dominates the production.\\
\indent In this paper, we focus on the low-energy positron population and on one likely source for these particles: the decay of radioactive species produced by the ongoing nucleosynthesis activity of our Galaxy, in particular $^{26}$Al, $^{56}$Ni and $^{44}$Ti. Recent works have shown that these elements could account for the annihilation signal observed by successive generations of gamma-ray telescopes \citep{Prantzos:2006,Higdon:2009}. In this work, we aim at providing an additional, alternative assessment of that possibility. We adapted the GALPROP public code for cosmic-ray propagation to simulate the transport and annihilation of radioactivity positrons in a model of our Galaxy. Using source spatial profiles based on typical distributions of massive stars and supernovae, we explored how the annihilation intensity distributions vary upon different prescriptions for the transport. The predicted emissions were then compared to the presently available observations, coming mostly from the INTEGRAL/SPI instrument.\\
\indent We start by recalling the main points of the physics of positron annihilation and transport in Sect. \ref{phys}, and by summarizing the main observational facts about Galactic low-energy positrons in Sect. \ref{obs}. In Sect. \ref{src}, we present the characteristics of $^{26}$Al, $^{56}$Ni and $^{44}$Ti in terms of contribution to the Galactic population of positrons. We introduce in Sect. \ref{model} the code used to model the propagation and annihilation of positrons in the Galaxy. Then, we present the simulated cases in Sect. \ref{simu} and discuss their results in Sect. \ref{results}.

\section{Positron physics}
\label{phys}

\subsection{Positron annihilation}
\label{phys_anni}

\indent Galactic low-energy positrons are associated with the strong 511\,keV celestial signal. For several reasons that will be reviewed in Sect. \ref{obs}, they are believed to be injected in the Galaxy with initial energies below a few MeV. On the other hand, low-energy positrons are created non-thermal in all the potential source processes, with mean initial energies of a few 100\,keV at least. To annihilate, however, most will have to be slowed down to near-thermal energies, at least below $\sim$100\,eV. This energy loss proceeds mostly through ionisation/excitation and Coulomb interactions in the interstellar gas, and it can take $\sim$10$^{5}$\,yr for a typical interstellar density of 1\,cm$^{-3}$ \citep{Jean:2006}. Over that time positrons travel at relativistic speeds, which opens the possibility for large-scale transport (see below).\\ 
\indent There is a variety of processes by which positrons can annihilate with electrons: direct annihilation with free or orbital electrons, formation of positronium by radiative recombination with free electrons or by charge-exchange with atoms, ...etc \citep[for a complete review, see][]{Guessoum:2005}. The physical and chemical properties of the medium where the annihilation takes place set the dominant annihilation channels and as a result, each ISM phase has a typical annihilation spectrum. As we will see later, most low-energy positrons seem to annihilate through the formation of a positronium atom (Ps), which is a short-lived bound state between a positron and an electron. This can occur in two different ways: either by charge-exchange, when non-thermalised positrons with typical kinetic energy of $\sim$10-100\,eV rip off electrons from H or He atoms; or by radiative recombination, when thermalised positrons combine with thermalised electrons. The Ps atoms occur in 75\% of the cases as ortho-Ps (parallel particle spins), which decays over $\mu$s time scales into 3 photons producing a continuum below 511\,keV, and in 25\% of the cases as para-Ps (anti-parallel particle spins), which decays over ns time scales into 2 photons with 511\,keV energies (in the centre-of-mass frame).\\
\indent Apart from the processes mentioned above, positrons can also annihilate directly with electrons at energies higher than $\sim$100\,eV and give rise to a Doppler-broadened line extending from $m_{e}c^{2}/2$ to $E_{k}+m_{e}c^{2}/2$, where $E_{k}$ is the positron kinetic energy and $m_{e}$ the electron/positron mass. This channel called in-flight annihilation is negligible for positron energies of $\leq$1\,MeV, but it becomes increasingly important at higher and higher energies. In typical ISM conditions, in-flight annihilation occurs for a maximum of $\sim$20-30\% of positrons with initial energies $\geq$10\,GeV \citep[this proportion decreases for stronger magnetic or radiation fields, see][]{Sizun:2006,Chernyshov:2010}.

\subsection{Positron transport}
\label{phys_trans}

\indent Once injected in the interstellar medium (ISM), relativistic positrons can propagate away from their sources in two ways, which we will term ballistic and collisionless transport. In the former case, positrons simply follow helicoidal trajectories along magnetic field lines and they experience repeated interactions with gas particles; in the process, positrons progressively lose their energy but experience little deviations\footnote{See Sect. 3 of \citet{Jean:2009}, the pitch angle of the particles remains nearly constant down to $\sim$10\,keV, and the propagation perpendicular to the field lines is negligible}. In the latter case, positrons are additionally scattered by magneto-hydrodynamic (MHD) perturbations associated with the interstellar turbulence and they random walk along and across the field lines, likely with different properties in the parallel and perpendicular directions; the process can be quite efficient, provided there are adequate MHD waves with which positrons can interact.\\
\indent From recent theoretical developments, it seems that interstellar MHD turbulence can be decomposed into Alfvenic, slow and fast magnetosonic modes. These modes are thought to be injected at spatial scales of the order of $\sim10^{20}-10^{21}$\,cm, as a result of differential rotation of the Galactic disk, superbubbles, and stellar winds and explosions. The turbulent energy is redistributed to smaller spatial scales through the interaction of wave packets in a so-called turbulence cascade. In the process, MHD modes can suffer various damping processes that could halt the cascade, thus preventing the propagation of MHD modes to smaller wavelengths.\\
\indent The properties of the turbulence cascade depend on the mode, and this has implications for the collisionless transport of energetic particles. The so-called Kolmogorov scaling applies to the Alfvenic and slow magnetosonic modes \citep{Lithwick:2001,Cho:2002}. In that case, however, the turbulent energy is preferentially redistributed perpendicularly to the magnetic field, which leads to an inefficient scattering of relativistic particles. Conversely, the fast magnetosonic modes follow the so-called Kraichnan scaling in an isotropic cascade \citep{Cho:2002,Cho:2003}, and they were shown to have the dominant contribution to the scattering of cosmic rays in the ISM \citep{Yan:2004}.\\
\indent \citet[][hereafter JGMF09]{Jean:2009} have made an extensive study of both the ballistic and collisionless transport modes for positron kinetic energies $\leq$10\,MeV, and we briefly summarise below their findings about collisionless transport. The authors focused on wave-particle resonant interactions occurring when the gyroradius of the positron is of order of the parallel wavelength of the MHD modes. In $\mu$G fields, this corresponds to scales of of $\sim10^{9}-10^{10}$\,cm for 1-10\,MeV particles, more than 10 orders of magnitude smaller than the scales at which turbulence is injected.\\
\indent A minimum wavelength for the MHD modes of the turbulence cascade is set by Landau damping occurring as the mode frequency approaches the cyclotron frequency of thermal protons. This corresponds to $\sim10^{8}-10^{9}$\,cm scales in most ISM phases and defines a minimum energy for particle-wave resonant interactions. In the case of positrons, this minimum energy is of order of $\sim$10-100\,keV, except in hot media with high magnetic field intensities, where it can exceed 1\,MeV. Yet, damping processes can cut off the turbulent cascade at higher wavelengths if the damping rate exceeds the energy transfer rate, and this is what is thought to happen for some ISM conditions. \citetalias{Jean:2009} determined that in the mostly neutral phases of the ISM, ion-neutral collisions halt the cascade at spatial scales of order $\sim10^{16}-10^{18}$\,cm for Alfven waves and $\sim10^{17}-10^{19}$\,cm for fast magnetosonic waves in warm and cold atomic phases, and at scales of order $\sim10^{17}-10^{20}$\,cm for both waves in molecular phases (where larger scales correspond to smaller gas densities for the atomic phases, and to larger cloud sizes for the molecular phases). In the ionised phases of the ISM, the Alfven wave cascade can proceed undamped down to the minimum spatial scales in both the hot and warm media, while the fast modes experience significant viscous damping at scales $\sim10^{13}-10^{14}$\,cm in the warm medium, and strong Landau damping at scales $\sim10^{17}-10^{18}$\,cm in the hot medium\footnote{Except maybe for quasi-parallel waves, if wave propagation angles are not efficiently randomised in wave-wave interactions}.\\
\indent The above results indicate that the turbulence cascade is quenched at spatial scales orders of magnitude greater than those required for resonant interactions of 1-10\,MeV positrons, except for Alfven waves in ionised media\footnote{Note that \citet{Ptuskin:2006} investigated turbulence dissipation through resonant interactions of cosmic rays, and concluded that this process may quench the Kraichnan turbulence cascade of fast modes at scales $\sim10^{13}$\,cm. While this was invoked as an explanation for the $\sim$1\,GeV/nucleon peak in the ratios of secondary-to-primary nuclei observed in the local cosmic rays, it may also be relevant for the transport of lower-energy particles.}. Yet, \citetalias{Jean:2009} argue that the scattering by Alfven waves is very likely inefficient due to the strong anisotropy of the associated turbulence. The alternative of Cerenkov resonance with fast modes, in which no precise gyroradius is required for the interaction, could potentially take place but under very restrictive conditions involving quasi-parallel fast waves and nearly perpendicular positron motion. On the other hand, non-resonant interactions with fast modes were proved to be quite efficient at scattering sub-MeV electrons in the solar wind \citep{Ragot:2006}, and a similar process may well operate for low-energy positrons in the ISM. Another option for an efficient collisionless transport of low-energy positrons is the injection of turbulence directly at the relevant small spatial scales, for instance through the streaming instability but more generally by any kind of fluid or kinetic instability. \citetalias{Jean:2009} concluded that positron scattering off cosmic-ray-driven waves is inefficient, but that scattering off self-generated waves would deserve a detailed investigation. Overall, \citetalias{Jean:2009} have presented several arguments against the scenario of collisionless transport of low-energy positrons, but the issue cannot yet be considered as definitely settled. Independently of \citetalias{Jean:2009}, \citet[][hereafter HLR09]{Higdon:2009} also came to the conclusion that the predominantly neutral phases of the ISM do not host the small-scale turbulence required for the resonant scattering of $\sim$MeV positrons. Yet, these authors argued from observations of particle propagation in the interplanetary medium that $\sim$MeV positrons should diffuse on MHD fluctuations in the ionised phases of the ISM, even if the exact nature of the process remains unknown.

\section{Positron observations}
\label{obs}

\subsection{Constraints on the annihilation sites}
\label{obs_anni}

\begin{figure}[!t]
\begin{center}
\includegraphics[width=\columnwidth]{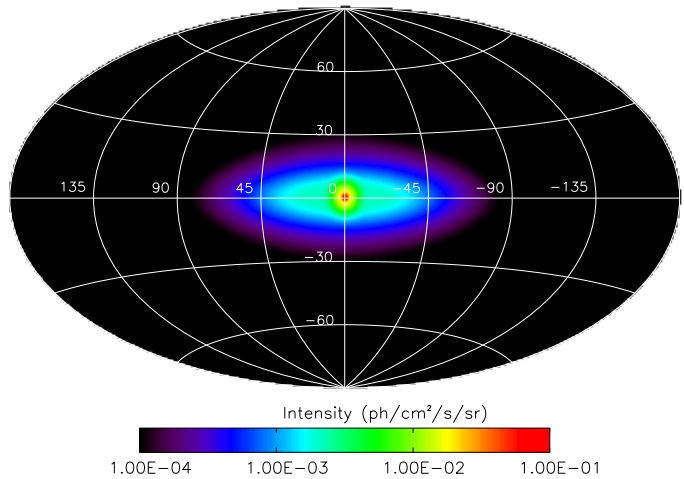}
\caption{Intensity distribution of the Galactic 511\,keV emission obtained by model-fitting to about 7 years of INTEGRAL/SPI observations, using the parameterized components adopted in \citet{Weidenspointner:2008}.}
\label{fig_obs_skymap}
\end{center}
\end{figure}

\indent In trying to identify the origin of low-energy positrons, the mapping of the annihilation emission by successive generations of balloon- and space-borne gamma-ray instruments provided a valuable piece of information. The most accurate picture available today is provided by the SPI telescope onboard the INTEGRAL satellite, which has been in orbit for about 9 years and is planned to operate at least until 2014. More than 100\,Ms of allsky observations concentrated mostly on the Galactic disk and bulge are now available. From this, it is clear that the morphology of the annihilation line emission is dominated by a strong central emission consisting of a very peaked part (FWHM of 2-3\deg) at the Galactic centre surrounded by a wider contribution (FWHM of 8-10\deg) from an outer bulge or halo (see Fig. \ref{fig_obs_skymap}). This inner emission comes on top of a fainter disk-like component extending up to at least $|l| < 50\deg$. The most interesting feature of this distribution is the relatively high bulge-to-disk (B/D) ratio that is obtained for the 511\,keV luminosity. Fitting projected 3D spatial distributions to the INTEGRAL/SPI data, luminosities of 1.2 / 3.1 $\times 10^{43}$ e$^{+}$/s for the bulge, and of 0.8 / 0.5 $\times 10^{43}$ e$^{+}$/s for the disk were inferred by \citet{Weidenspointner:2008a} for two sets of equally-likely models\footnote{The values are also based on the assumption of a positronium fraction of 0.967, as determined by \citet{Jean:2006} from the spectral analysis of the annihilation emission.}. With values of order 2-6, the B/D luminosity ratio is higher than those obtained for the distributions of classical astrophysical objects or interstellar gas (with the caveat, however, that the outer disk emission is poorly constrained; see Sect. \ref{results_integral}).\\
\indent Valuable information about where exactly positrons end their lives could be obtained from the spectral analysis the 511\,keV line, permitted by the high energy resolution of the SPI instrument. The annihilation emission observed with SPI in the inner Galactic regions, the 511\,keV line and the continuum below down to $\sim$400\,keV, indicates that most positrons annihilate in the warm medium of the ISM \citep{Churazov:2005,Jean:2006}. Approximately one half annihilate in the warm ionised phase, through the formation of Ps by radiative recombination. The other half of the positrons annihilate in the cold and warm neutral phases shortly before thermalisation, through the formation of Ps by charge-exchange. Most of the positrons that feed the 511\,keV line emission therefore annihilate through the formation of a Ps state. The so-called positronium fraction $f_{Ps}$ inferred from SPI is quite high, of the order of $\geq$95\%, and the remaining few \% correspond to direct annihilation with free electrons in the warm ionised medium.

\subsection{Constraints on the injection energy}
\label{obs_inj}

\indent We now turn to constraints on the origin of positrons, and in particular on their initial energy and on the reasons why the positrons involved in the 511\,keV line emission are generally considered to be low-energy particles. For a given positron population, the level of in-flight radiation at any gamma-ray energy $\geq m_{e}c^{2}/2$ increases with the initial energy of the particles, which allows to get constraints on this parameter from observations in the 1-100\,MeV range. Under the assumption that positrons remain confined to the Galactic bulge regions from their injection to their annihilation, \citet{Sizun:2006} and \citet{Beacom:2006} concluded that the positrons responsible for the 511\,keV emission must be injected with initial energies below a few MeV otherwise their in-flight annihilation would give rise to an emission excess from the bulge, in contradiction with the CGRO/COMPTEL observations in the 1-30\,MeV range. A less stringent constraint of a few GeV can be obtained assuming an injection of positrons $\sim 10^{5}$\,yrs ago by a non-stationary source process, followed by a slowing-down in a magnetic field of a few 100\,$\mu$G in the inner Galactic regions \citep{Chernyshov:2010}. We will however focus on a stationary source in this work.\\
\indent Constraints on the initial energy of positrons annihilating in the bulge could also be obtained from other radiations than in-flight annihilation. Positrons with energies $\geq 100$\,MeV would emit through inverse-Compton scattering, Bremsstrahlung, and synchrotron, and add their contribution to the gamma-ray and radio emission from cosmic rays interacting with the ISM \citep{Strong:2007,Strong:2011}. Yet, the production rate of positrons annihilating at low energy very likely outnumbers that of cosmic-ray electrons/positrons. In the model of Galactic cosmic rays used in \citet{Strong:2010}, the injection rate of primary electrons with energies 100\,MeV-100\,GeV is $\sim$ 10$^{42}$ e$^{-}$/s; adding secondary electrons/positrons created by inelastic collisions of cosmic ray nuclei with interstellar gas approximately doubles the rate \citep{Porter:2008}. As seen previously, however, the positron annihilation rate inferred from INTEGRAL/SPI observations is a few times $10^{43}$ e$^{+}$/s; so if these positrons were to have initial energies $\geq 100$\,MeV, they would certainly dominate the Galactic high-energy or radio emission, at least over certain energy ranges. Although current models of the Galactic high-energy and radio emission can probably accommodate a population of high-energy positrons in addition to conventional cosmic rays from supernova remnants, it seems improbable that this additional component have similar injection energies and a flux an order of magnitude higher. Moreover, the very different morphologies of the 511\,keV and $\sim$10\,MeV-100\,GeV skies do not seem to support a high initial energy for positrons, since they would first have to lose their high initial energies mostly in the disk and ultimately annihilate in a very narrow region in the Galactic bulge.

\section{Positron sources}
\label{src}

\indent Positrons are naturally created in the $\beta^+$-decay of some unstable proton-rich nuclei. The main species anticipated to significantly contribute to the positron production through $\beta^+$-decay are $^{26}$Al, $^{56}$Ni and $^{44}$Ti \citep[$^{22}$Na may also bring some contribution, but we will not consider it here; see][]{Prantzos:2004}. Nucleosynthesis is often taken as the most plausible origin for the Galactic positrons because the existence and decay of the above-mentioned species can be established from various experimental results. The most direct evidence comes from observations of the characteristic gamma-ray lines or fluorescence X-ray lines that accompany the decay of the above-mentioned isotopes \citep[see for instance][]{Renaud:2006,Leising:1990}. Then, indirect proof can be obtained from the interpretation of supernovae lightcurves, which are powered by the energetic decay products of $^{56}$Ni and $^{44}$Ti, or from the measurement of present-day cosmic abundances of the daughter nuclei, which result from the cumulated nucleosynthesis history of the Galaxy \citep{Motizuki:2004,The:2006}. The isotope production yields inferred from these various observations would actually be sufficient to provide after $\beta^+$-decay the estimated $\sim10^{43}$ e$^{+}$/s that power the diffuse annihilation emission observed by INTEGRAL. Yet, as we will see below, these yields cannot be straightforwardly translated into a Galactic positron injection rate because all positrons do not necessarily escape the production sites of their parent radio-isotopes.\\
\indent In the following, we present the properties of the positrons produced by the decay of the three above-mentioned radionuclides: the plausible large-scale spatial distribution of their injection sites, the spectrum they have when they are released into the ISM, and their production rate inferred from the estimated nucleosynthesis yields for the parent isotopes. For a more detailed discussion about the subject, we refer the reader to \citet{Prantzos:2011}.

\subsection{The $^{26}$Al radio-isotope}
\label{src_26al}

\indent The $^{26}$Al isotope is believed to be produced predominantly by massive stars at various stages of their evolution, with typical total yields of a few 10$^{-4}$-10$^{-5}$\,M$_{\odot}$ per star \citep{Prantzos:1996}. It is released in the ISM by stellar winds and by core-collapse supernova explosions. The mean $^{26}$Al lifetime of about 1\,Myr allows it to escape its production site and diffuse $\sim$10-100\,pc away before decaying into $^{26}$Mg. The transport of $^{26}$Al away from its stellar sources imply that the majority of $^{26}$Al positrons are released directly into the ISM and do not suffer energy losses prior to their injection.\\
\indent The distribution of $^{26}$Al in the Galaxy is truly diffuse and follows that of massive stars, as confirmed by the mapping of the 1809\,keV decay emission \citep{Pluschke:2001} and its correlation with the microwave free-free emission from HII regions or the infrared emission from heated dust \citep{Knodlseder:1999}. The Galactic mass distribution of $^{26}$Al inferred from the 1809\,keV emission is strongly concentrated at galactocentric radii 3 to 6\,kpc, at the position of the so-called molecular ring \citep{Martin:2009a}. The injection sites of $^{26}$Al positrons are thus expected to be distributed in an annular disk and along spiral arms \citep[such as the thin disk and spiral arms components that represent the contribution of localised HII regions surrounding massive stars in the NE2001 model for the Galactic distribution of free electrons by][]{Cordes:2002}.\\
\indent Because of the long lifetime of $^{26}$Al, positrons are very likely injected into the ISM with their original, unaltered $\beta$-spectrum. The latter is described by the formula given in \citet[][hereafter CL93]{Chan:1993} and has a mean of $\sim$0.4\,MeV and a maximum of 1.2\,MeV. We note here that most $^{26}$Al positrons are very likely released in the hot, tenuous, and highly turbulent interiors of the superbubbles blown by massive star clusters. In these conditions, relativistic positrons may well experience stochastic, second-order Fermi acceleration in the strong MHD turbulence, or even regular, first-order Fermi acceleration in the many primary/secondary shocks that travel the bubbles. This could significantly modify the initial $\beta$-spectrum and bring positrons to higher energies, which could increase their range and move their annihilation sites further. This effect is, however, beyond the scope the present paper and would deserve a dedicated study.\\
\indent The total injection rate of $^{26}$Al positrons can be computed from the estimated total stationary $^{26}$Al mass of 1.7-2.7\,M$_{\odot}$ in the Galaxy \citep{Martin:2009a,Wang:2009}, where the uncertainty on the total Galactic $^{26}$Al mass arises from uncertainties on the exact spatial distribution of the isotope in the Galaxy. Taking into account  a $\beta^+$-decay branching ratio of 0.82, this translates into a $^{26}$Al positron injection rate of $(0.20-0.31) \times 10^{43}$ e$^{+}$/s.

\subsection{The $^{44}$Ti radio-isotope}
\label{src_44ti}

\indent The $^{44}$Ti isotope is synthesised by explosive Si-burning deep in the stellar ejecta during core-collapse and thermonuclear supernova explosions (ccSNe and SNe Ia respectively), with typical yields of a few 10$^{-4}$-10$^{-5}$\,M$_{\odot}$ per event. The mean $^{44}$Ti lifetime is about 85\,yr, which implies that the radio-isotope remains trapped in the young supernova remnant until it decays into $^{44}$Sc and then $^{44}$Ca. Most $^{44}$Ti positrons are therefore released in the envelope of the exploded star and their further transfer to the ISM depends on the unknown transport conditions in the ejecta and on the explosion properties. This must be taken into account when translating $^{44}$Ti yields into positron injection rates (see below).\\
\indent The injection sites of $^{44}$Ti positrons sample the time-averaged distribution of supernova explosions in our Galaxy. Core-collapse supernovae proceed from massive stars and are thus distributed like them, in a thin annular disk and along spiral arms (see \ref{src_26al}). Thermonuclear supernovae have a slightly more subtle evolutionary pathway and actually depend on both the old and young stellar populations. \citet{Sullivan:2006} showed from observations of external galaxies that the rate of SNe Ia depends on both the star formation rate and the total stellar mass\footnote{A scenario based only on stellar mass, that is on the older and lower-mass star population, is ruled out at $>$99\% confidence level.}. The spatial distribution of SNe Ia is therefore a combination of that of massive stars and that of stellar mass. In our Galaxy, most of the stellar mass is distributed in an exponential disk with a central hole and in a ellipsoidal bulge, with approximately the same mass in each component: 2.15 and 2.03 $\times$ 10$^{10}$\,M$_{\odot}$, respectively \citep[see the determination of the three-dimensional shapes and parameters in][]{Robin:2003}. Overall, the source distribution for $^{44}$Ti positrons has three components: a thin annular disk (for ccSNe and prompt SNe Ia), and an exponential disk with central hole and a bulge (for delayed SNe Ia). Further down, we quantify the relative contribution of each component to the total source term.\\
\indent Because of the intermediate lifetime of $^{44}$Ti, the population of decay positrons entering the ISM may be affected by the escape from the stellar ejecta. \citet{Martin:2010} computed the escape or survival fractions of $^{44}$Ti positrons for two extreme transport modes in the ejecta: either positrons can freely stream through the stellar envelope, or they are trapped at their birth place by some strong magnetic turbulence. In the free streaming case, the escape fractions are quite high, ranging from 97\% for light 2\,M$_{\odot}$ ejecta to 83\% for more massive 14\,M$_{\odot}$ ejecta; in the trapped case, the survival fractions are lower, ranging from 91\% to 36\% for the same mass range (assuming in both cases a typical 10$^{51}$\,erg explosion kinetic energy)\footnote{These estimates for the escape fractions are consistent with the results obtained by \citetalias{Chan:1993}, except for the 14\,M$_{\odot}$ ejecta mass, for which they found a survival fraction of 8\% only in the trapped case. The origin of the discrepancy may lie in the approximation they used in their calculation, while \citet{Martin:2010} performed the complete integration over time and energy. Another difference between both studies is that \citetalias{Chan:1993} added a so-called slow positron survival fraction for positrons that are thermalised but do not annihilate in the continuously-thinning ejecta. Yet, in the ejecta assumed to be neutral, the dominant annihilation process is positronium formation, which occurs over comparatively very short time scales once positrons have been slowed down below $\sim$100\,eV. So if neutral atoms are indeed the dominant species in the rapidly-cooling ejecta, no more than a few \% of the thermalised positrons are expected to survive.}. Although $^{44}$Ti positrons do suffer energy losses on their way out of the stellar ejecta, the mean energy of the escaping/surviving population remains close to the initial one, whatever the transport mode. So most $^{44}$Ti positrons are injected into the ISM with a spectrum close to the original, unaltered $\beta$-spectrum, which has a mean of $\sim$0.6\,MeV and a maximum of 1.5\,MeV. We note, however, that \citet{Zirakashvili:2011} considered the possibility that positrons from the $^{44}$Ti decay chain constitute a pool of mildly relativistic particles that can be accelerated to ultrarelativistic energies in the remnants of supernova explosions of all types. In their model, the $\sim$1\,MeV positrons from $^{44}$Sc decay undergo stochastic pre-acceleration up to $\sim$100\,MeV in the turbulent upstream region of the reverse shock, and are then further accelerated up to multi-TeV energies by the diffusive shock acceleration mechanism at the shock. This effect is, however, beyond the scope the present paper.\\
\indent The total injection rate of $^{44}$Ti positrons can be computed from the frequencies of ccSNe and SNe Ia, and the $^{44}$Ti yields and positron escape fractions for each type of event. From \citet{Tammann:1994}, ccSNe are estimated to occur at a rate $\sim$2.1 ccSNe/century while SNe Ia occur at a rate $\sim$0.4 SNe Ia/century. Using the empirical relation from \citet{Sullivan:2006}, on which we based the spatial distribution of SNe Ia, the occurrence rate of SNe Ia is
\begin{align}
R_{\textrm{SNIa}}= \quad & M_{\star} \times (5.3 \pm1.1) . 10^{-14} \, \textrm{yr}^{-1} \textrm{M}_{\odot}^{-1} \notag \\
+ & SFR \times (3.9 \pm0.7) . 10^{-4} \, \textrm{yr}^{-1} (\textrm{M}_{\odot} \, \textrm{yr}^{-1})^{-1}
\end{align}
we also find a rate of $\sim$0.4 SNe Ia/century if we take a total star formation rate $SFR=$ 4\,M$_{\odot}$\,yr$^{-1}$ \citep{Diehl:2006} and the above-mentioned total stellar mass $M_{\star}=$ (2.15+2.03)$\times$ 10$^{10}$\,M$_{\odot}$ \citep{Robin:2003}. Then, a star formation rate of 4\,M$_{\odot}$\,yr$^{-1}$ corresponds to $\sim$2.1 ccSNe/century for a typical initial mass function \citep[see Table 1 in][]{Diehl:2006}. Regarding the $^{44}$Ti yields of supernovae, observational estimates are very scarce \citep[two direct detections, Cassiopeia A and G1.9+0.3, and one indirect, SN1987A; see][]{Renaud:2006,Borkowski:2010,Motizuki:2004}, so it is risky to compute some average Galactic $^{44}$Ti production rate by integrating theoretical yields over a range of supernova progenitors. Instead, we simply assumed that ccSNe and SNe Ia eject on average 2.0 $\times 10^{-4}$\,M$_{\odot}$ and 2.0 $\times 10^{-5}$\,M$_{\odot}$ per event, respectively. These values agree with the compilation of observational constraints done in \citet{Martin:2010} and with the recently-estimated range for G1.9+0.3 \citep{Borkowski:2010}. Overall, this corresponds to a mean Galactic $^{44}$Ti production rate of 4.2 $\times 10^{-6}$\,M$_{\odot}$\,yr$^{-1}$, roughly consistent with the estimate of 5.5 $\times 10^{-6}$\,M$_{\odot}$\,yr$^{-1}$ by \citet{The:2006} based on the present-day solar $^{44}$Ca abundance and a Galactic chemical evolution model. This $^{44}$Ti production rate can be translated into a positron production rate by applying a factor 0.94 for the $\beta^+$-decay branching ratio of the decay chain and using an escape fraction of 100\%. In the absence of strong observational constraints, the latter value was chosen because the escape fraction in models is $\geq$50\% for most ejecta masses, explosion energies, and transport conditions, so using 100\% gives the correct order of magnitude and allows an easy scaling of the results. Eventually, the $^{44}$Ti positron injection rate is $0.34 \times 10^{43}$ e$^{+}$/s (with only $\sim$2\% being contributed by SNe Ia), and we consider an uncertainty range of $\pm$50\% due to the uncertainty on the Galactic $^{44}$Ti production rate\footnote{In \citet{The:2006}, the authors found a $\sim2$ times larger uncertainty range for the Galactic $^{44}$Ti production rate due to unknown parameters in the Galactic chemical evolution model; they also note that the entire range may be shifted to lower values if there is a nucleosynthesis channel to produce $^{44}$Ca directly.}.

\subsection{The $^{56}$Ni radio-isotope}
\label{src_56ni}

\indent The $^{56}$Ni isotope is synthesised by explosive Si-burning deep in the stellar ejecta during core-collapse and thermonuclear supernova explosions, with typical yields of the order of 10$^{-1}$\,M$_{\odot}$ per event. The characteristic time of the decay chain to $^{56}$Co and then to $^{56}$Fe is $<$1\,yr, which implies that all positrons are released in the stellar ejecta at the late supernova / early remnant phase. This makes escape a critical point, which actually counterbalances the large $^{56}$Ni yields.\\
\indent Although $^{56}$Ni is produced by both ccSNe and SNe Ia, we will see below that SNe Ia are expected to dominate over \mbox{ccSNe} in terms of contribution to the Galactic positron population through that isotope. Consequently, the injection sites of $^{56}$Ni positrons follow the time-averaged distribution of SNe Ia in our Galaxy. The source distribution has three components like that of $^{44}$Ti positrons - a thin annular disk, an exponential disk, and a bulge - but the relative contribution of each component to the total positron production rate is different, as we will see below.\\
\indent Because of the short timescale of the $^{56}$Ni decay chain, the population of decay positrons eventually entering the ISM is strongly affected by the travel through the stellar ejecta. The escape of $^{56}$Ni positrons is favored by ejecta mixing, which lifts iron up to the lower-density ejecta surface, and by low confinement, which decreases the column density experienced by positrons in the ejecta. Models exploring these effects result in escape fractions up to $\sim$10\% for SNe Ia, while escape fractions for ccSNe reach a few \% only for the light and rare type Ib/Ic explosions and in the limiting case of a fully mixed ejecta \citepalias[see for instance][]{Chan:1993}. The comparison of such predictions with the late lightcurves of SNe Ia indicates an escape fraction in the range $\sim$2-6\% \citep[][but see also the warning of \citet{Lair:2006}]{Milne:1999}. Alternatively, \citet{Martin:2010} translated the non-detection of 511\,keV emission from the youngest and most nearby supernova remnants into an upper limit on the escape fraction of 12\% for SNe Ia. Overall, SNe Ia are thought to be the dominant source of $^{56}$Ni positrons when compared to ccSNe: their lower occurrence rate (0.4 versus 2.1 SNe/century) is compensated by their higher iron yield per event (0.6 versus 0.1\,M$_{\odot}$, typically), and their average positron escape fraction is very likely one order of magnitude higher. Due to the strong energy losses experienced in the fresh and dense ejecta of SNe Ia, escaping/surviving $^{56}$Ni positrons very likely have a spectral distribution that differs from the original $\beta$-spectrum. Depending on the actual mixing and density profile of the ejecta, the mean kinetic energy of the positron population may be shifted from $\sim$0.6\,MeV at decay to $\sim$0.3\,MeV or even below when entering the ISM \citepalias[see the Fig. 4 of][]{Chan:1993}. We will discuss later on the impact of this effect on the predicted annihilation emission.\\
\indent With the previously-adopted SNIa occurrence rate of 0.4 per century and a typical $^{56}$Ni yields of 0.6\,M$_{\odot}$, the total positron injection rate can be computed by taking into account a factor of 0.19 for the $\beta^+$-decay branching ratio of the decay chain and assuming an escape fraction of 5\%. This corresponds to a $^{56}$Ni positron injection rate is $1.53 \times 10^{43}$ e$^{+}$/s. Yet, the positron escape fraction is a crucial parameter that is not strongly constrained by observations. We therefore considered an uncertainty range of one order of magnitude for this factor, for a range of 1 to 10\%. This translates into an uncertainty range on the $^{56}$Ni positron injection rate of $(0.31-3.10) \times 10^{43}$ e$^{+}$/s.
\begin{table*}[!ht]
\begin{minipage}[][9.8cm][c]{\textwidth}
\begin{center}
\caption{Bulge, disk, and total annihilation radiation luminosities for $^{56}$Ni, $^{44}$Ti, and $^{26}$Al positrons in the three transport configurations tested. Also indicated are the corresponding bulge-to-disk ratios, and the fractions of positrons that annihilate in the Galaxy.}
\begin{tabular}{|c|c|c|c|c|c|c|}
\hline
\celltspace Source & Transport & Bulge (R $\leq$ 3\,kpc) & Disk (R $>$ 3\,kpc) & Total Galaxy & Bulge/Disk ratio & Annihilation fraction \cellbspace \\
\hline 
\celltspace \multirow{3}{*}{$^{56}$Ni only} & A & 0.32 & 0.58 & 0.91 & 0.56 & 1.00 \\
\celltspace & B & 0.16 & 0.47 & 0.63 & 0.35 & 0.68 \\
\celltspace & C & 0.09 & 0.39 & 0.48 & 0.22 & 0.57 \cellbspace \\
\hline
\celltspace \multirow{3}{*}{$^{44}$Ti only} & A & 0.01 & 0.18 & 0.20 & 0.08 & 1.00 \\
\celltspace & B & 0.01 & 0.11 & 0.12 & 0.10 & 0.63 \\
\celltspace & C & 0.01 & 0.10 & 0.11 & 0.06 & 0.59 \cellbspace \\
\hline
\celltspace \multirow{3}{*}{$^{26}$Al only} & A & 0.01 & 0.13 & 0.14 & 0.08 & 1.00 \\
\celltspace & B & 0.01 & 0.10 & 0.11 & 0.09 & 0.77 \\
\celltspace & C & 0.01 & 0.09 & 0.10 & 0.07 & 0.73 \cellbspace \\
\hline
\celltspace \multirow{3}{*}{$^{56}$Ni+$^{44}$Ti+$^{26}$Al} & A & 0.35 & 0.90 & 1.25 & 0.39 & 1.00 \\
\celltspace & B & 0.18 & 0.68 & 0.86 & 0.27 & 0.69 \\
\celltspace & C & 0.10 & 0.59 & 0.69 & 0.17 & 0.56 \cellbspace \\
\hline
\end{tabular}
\label{tab_lumi}
\end{center}
Note to the table: The luminosities in columns 3 to 5 are given in 10$^{43}$\,ph\,s$^{-1}$ for the mean positron injection rates determined in Sect. \ref{src}. They correspond to 511\,keV emission from parapositronium and direct annihilation, for a positronium fraction $f_{Ps}$=0.95.
\end{minipage}
\end{table*}

\section{The transport code}
\label{model}

\indent The propagation of positrons in the Galaxy was modelled with a modified version of the publicly-available GALPROP code (see http://galprop.stanford.edu). The code was originally created to analyse in a consistent way the growing body of data available on Galactic cosmic rays (direct particle measurements, diffuse emissions in radio and at high energies). Nevertheless, it is general enough to be adapted to other studies of particle transport in the Galaxy.\\
\indent In the following, we briefly review the constituents of the GALPROP model that are relevant to the study of nucleosynthesis positrons. We also introduce the modifications implemented to simulate the propagation and annihilation of positrons in the Galaxy.

\subsection{Transport processes}
\label{model_trans}

\begin{figure}[!t]
\vspace{1cm}
\begin{center}
\includegraphics[width=\columnwidth]{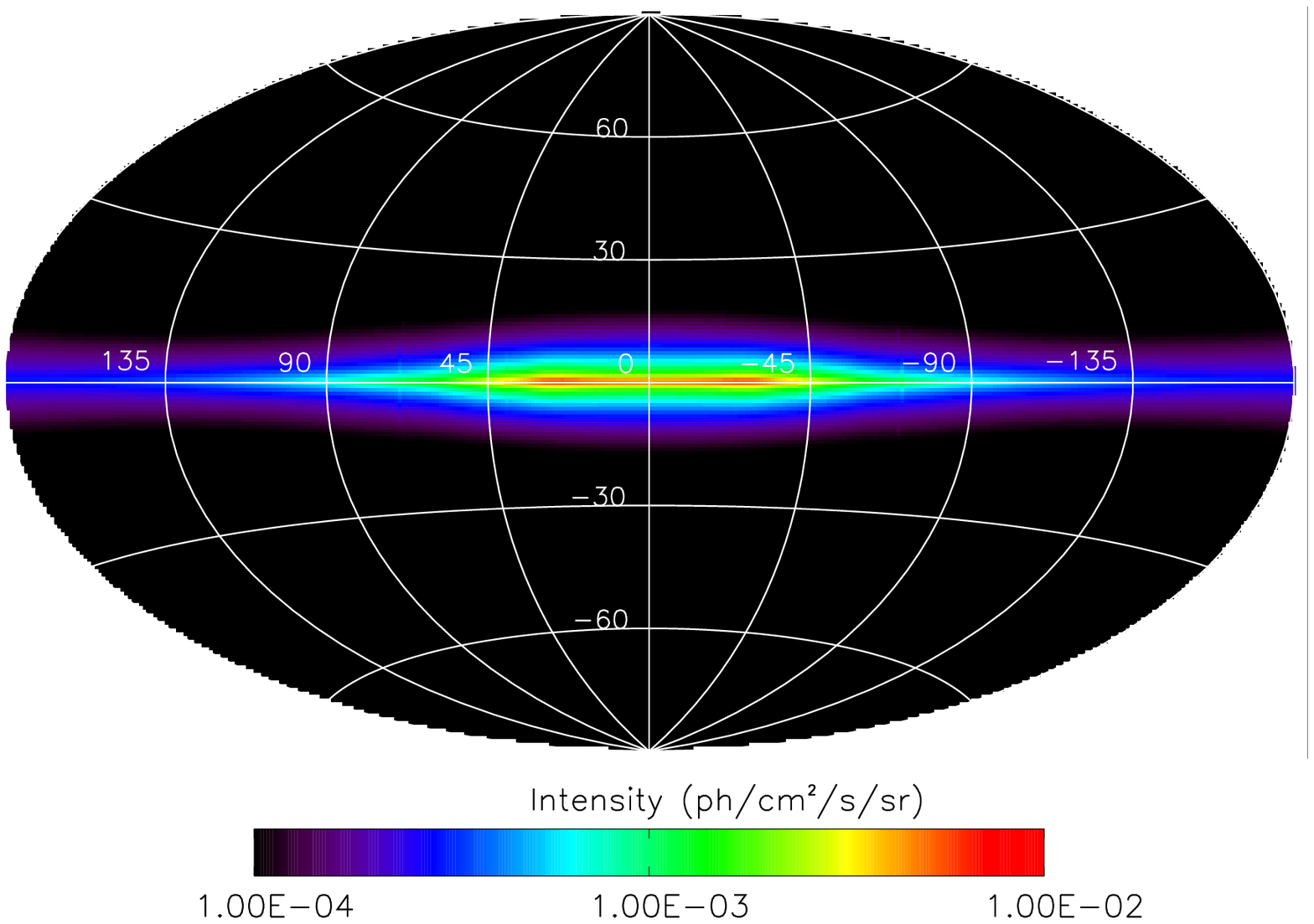}
\includegraphics[width=\columnwidth]{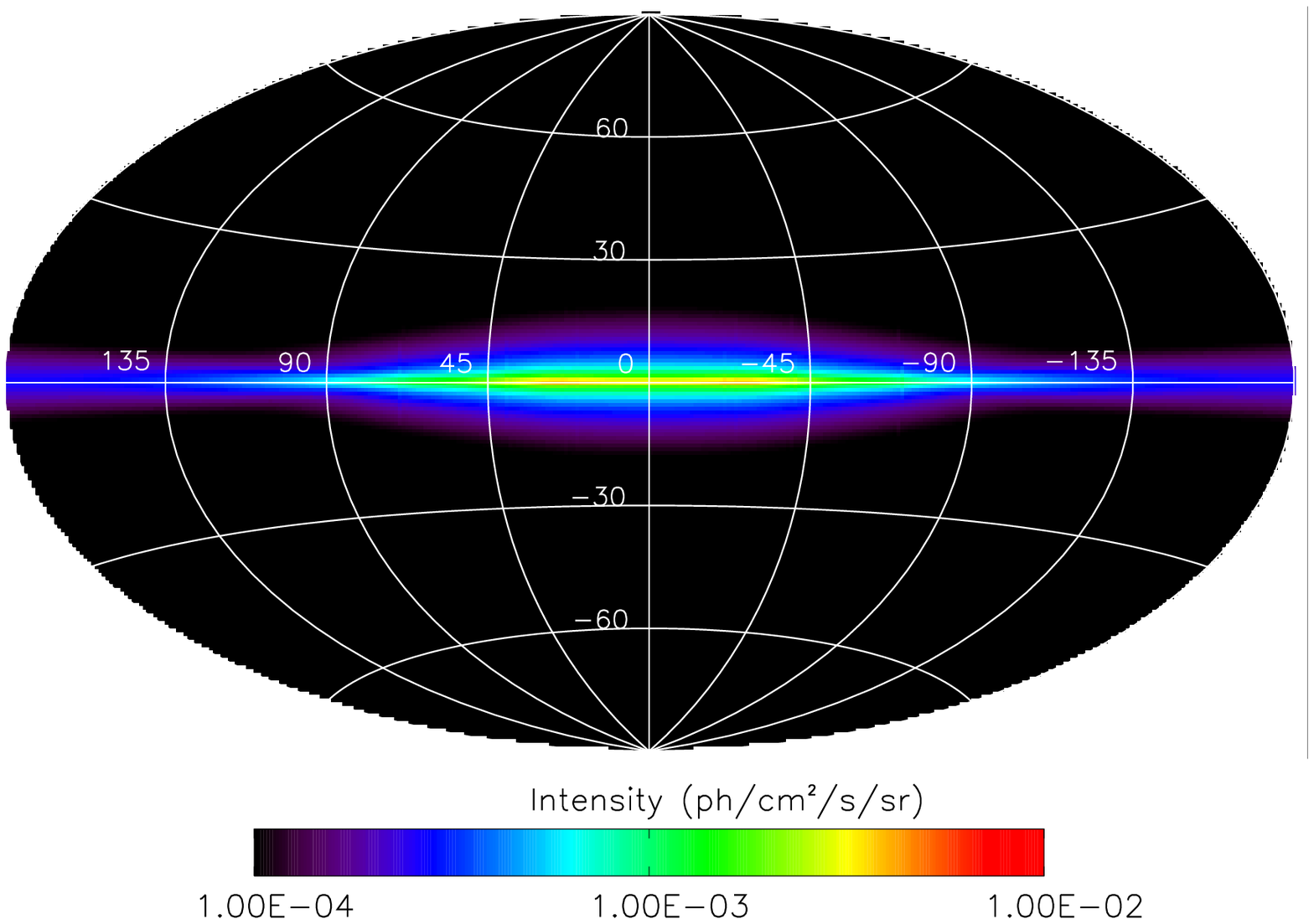}
\includegraphics[width=\columnwidth]{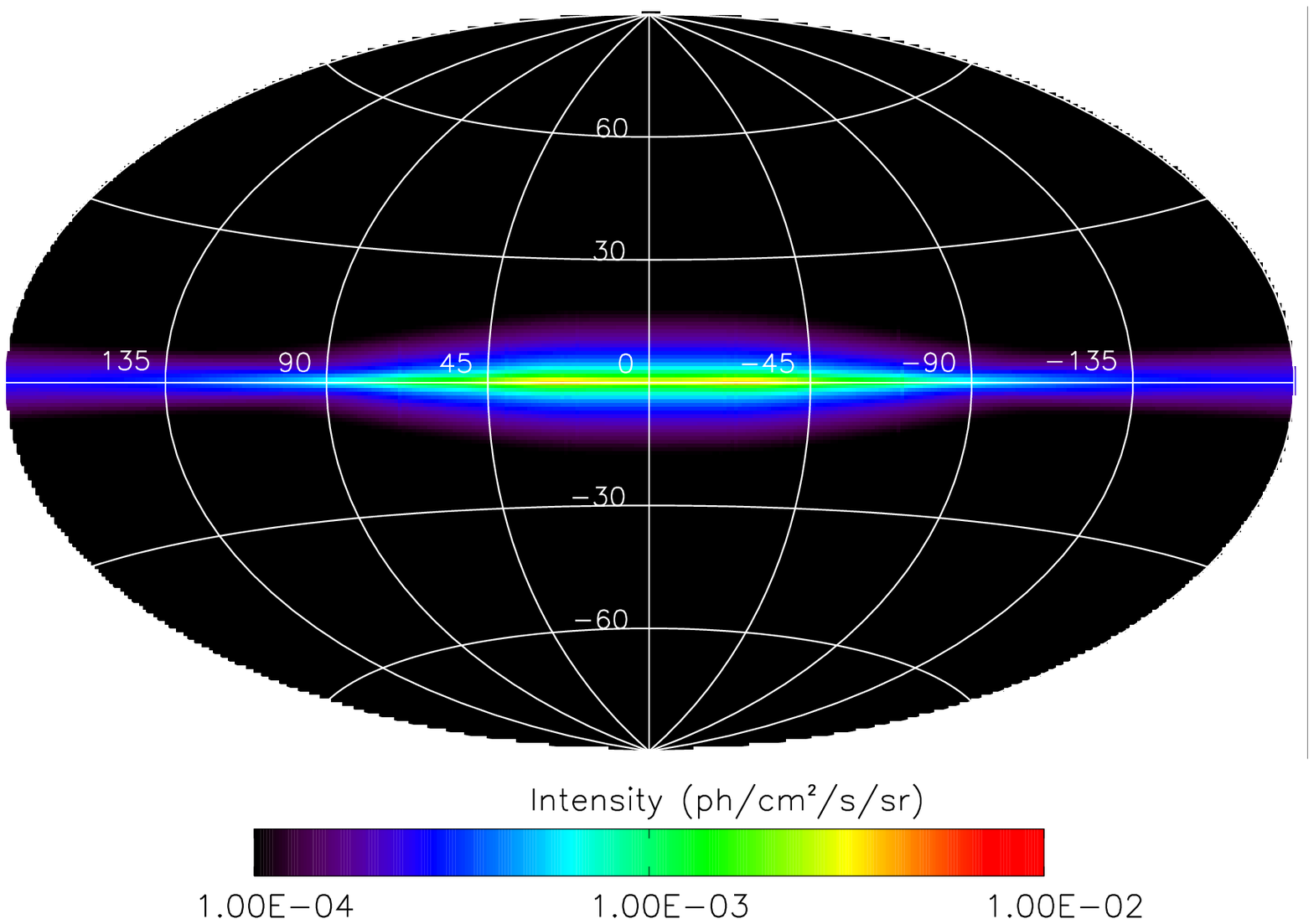}
\caption{Predicted 511\,keV intensity distributions for the annihilation of $^{26}$Al positrons in each transport configuration. From top to bottom, the transport was assumed to be collisionless everywhere (case A, small diffusion coefficient), collisionless in the bulge and ballistic out of it (case B, inhomogeneous case), and ballistic everywhere (case C, large diffusion coefficient). The given intensities correspond to parapositronium annihilation only and were normalised to a positron injection rate of 10$^{43}$ e$^{+}$/s, with a positronium fraction of 0.95.}
\label{fig_26Al_skymap}
\end{center}
\newpage
\end{figure}
\indent GALPROP numerically solves a general diffusion-convection equation for a given source distribution. The transport mechanisms included in the code are diffusion in position and momentum, convection away from the plane and the associated adiabatic cooling, momentum losses from many processes, radioactive decay, and nuclear fragmentation. The equation is solved using a Crank-Nicholson implicit second-order scheme, with free particle escape assumed at the spatial boundaries. A typical simulation is run for decreasing time steps until a steady state is achieved for all species over the entire computational grid. The present modelling of positron propagation in the Galaxy includes diffusion in position and energy losses, but no convection or diffusive reacceleration (and of course no nuclear process).\\
\indent By construction, the transport in position space is treated as a diffusive process and we will explain in Sect. \ref{simu_diff} how this can be justified for the cases we considered. We modified the code to allow the simulation of inhomogeneous diffusion (with different properties in the bulge and in the disk for instance). To avoid potential problems arising from discontinuities in the grid of diffusion coefficients, we implemented smooth transitions between regions with different diffusion properties. In the case of a more efficient positron diffusion in the disk compared to the bulge (see Sect. \ref{simu_diff}), the diffusion coefficient is described by a Gaussian-type function having its lowest value at the Galactic centre and reaching its highest value beyond a certain scale radius and height above the plane.\\
\indent Regarding the transport in momentum space, GALPROP includes the main energy loss processes for high-energy charged particles, such as Bremsstrahlung, inverse-Compton, and synchrotron, but the dominant one in the keV-MeV range are Coulomb interactions and ionisation/excitation of atoms. Ionisation/excitation are implemented following the prescription of \citet{Pages:1972}, based on Bethe's theoretical formula, with experimental values for the ionisation potentials of neutral H and He, and without correction for the density effect. Coulomb losses are implemented following the prescription of \citet{Ginzburg:1979} for the cold plasma limit. As noted below, we also implemented direct in-flight annihilation in the code. This constitutes an additional catastrophic loss process, but its contribution for MeV positrons is insignificant.\\
\indent Overall, the transport equation solved for positron propagation in the Galaxy (until a steady state is achieved) is the following
\begin{equation}
\frac{\partial \varphi}{\partial t}= \nabla \left( D \nabla \varphi \right) - \frac{\partial}{\partial E_{p}}\left( \varphi \dot{E}_{p} \right) + Q
\end{equation}
where $\varphi$ is the positron distribution function, $D$ is the spatial diffusion coefficient, $\dot{E}_{p}$ is the positron energy loss rate, and $Q$ is the source term. All these quantities depend on position $\vec{r}$ and positron energy $E_{p}$. The equation does not contain an explicit term for annihilation. As we will see below in Sect. \ref{model_anni}, the annihilation rate is computed at each position from the rate of positrons being slowed down below $\sim$100\,eV.

\subsection{Interstellar medium}
\label{model_ism}

\indent GALPROP includes average analytical spatial distributions for the main gas states: molecular (H$_2$), atomic (HI), and ionised (HII). These distributions are used in the computation of the propagation of cosmic-rays throughout the Galaxy, for the determination of energy losses for instance. Then, for the prediction of diffuse emissions from cosmic rays interacting with interstellar gas (such as Bremsstrahlung or $\pi^0$ production and decay), GALPROP allows to recover the fine spatial structure of the gas through the use of the observed gas column densities (in the HI 21\,cm and CO 2.6\,mm emission lines). In our simulations of the annihilation emission, however, we used only the analytical gas distributions because the observational constraints at our disposal, coming mostly from INTEGRAL/SPI, have an angular resolution several times above that of HI or CO surveys.\\
\indent Atomic hydrogen is the dominant gas phase in terms of total mass and has a relatively large filling factor. The GALPROP code uses a 2D analytical distribution for HI. The radial distribution is taken from \citet{Gordon:1976}, while the vertical distribution is from \citet{Dickey:1990} for 0 $\leq$ R $\leq$ 8\,kpc and \citet{Cox:1986} for R $\geq$ 10\,kpc, with linear interpolation between the two ranges.\\
\indent Molecular gas accounts for about a third of the total gas mass of the Milky Way, but it is concentrated in complexes of dense, massive clouds with low filling factors. The GALPROP code employs a 2D analytical distribution for H$_2$, using the model from \citet{Ferriere:2007} for R $\leq$ 1.5\,kpc, the model from \citet{Bronfman:1988} for 1.5\,kpc $<$ R $<$ 10\,kpc, and the model from \citet{Wouterloot:1990} for R $\geq$ 10\,kpc.\\ 
\indent The ionised hydrogen makes up about 10\% of the total gas mass of the Milky Way and is therefore the least massive component. Yet, It actually occupies most of the Galactic volume and has a very extended distribution across the plane. The GALPROP code uses a two-component model for the HII distribution: a thin disk tracing localised HII regions in the plane, mostly in the molecular ring and spiral arms, and a thick disk representing the more diffuse warm ionised gas that exists outside the well-defined HII regions. The former is modelled by the 2D/axisymetric thin disk component of the NE2001 model \citep{Cordes:2002}, while the latter is modelled by a vertical exponential distribution with the scale height determined by \citet{Gaensler:2008}.

\subsection{Annihilation}
\label{model_anni}

\begin{figure}[!ht]
\vspace{1cm}
\begin{center}
\includegraphics[width=\columnwidth]{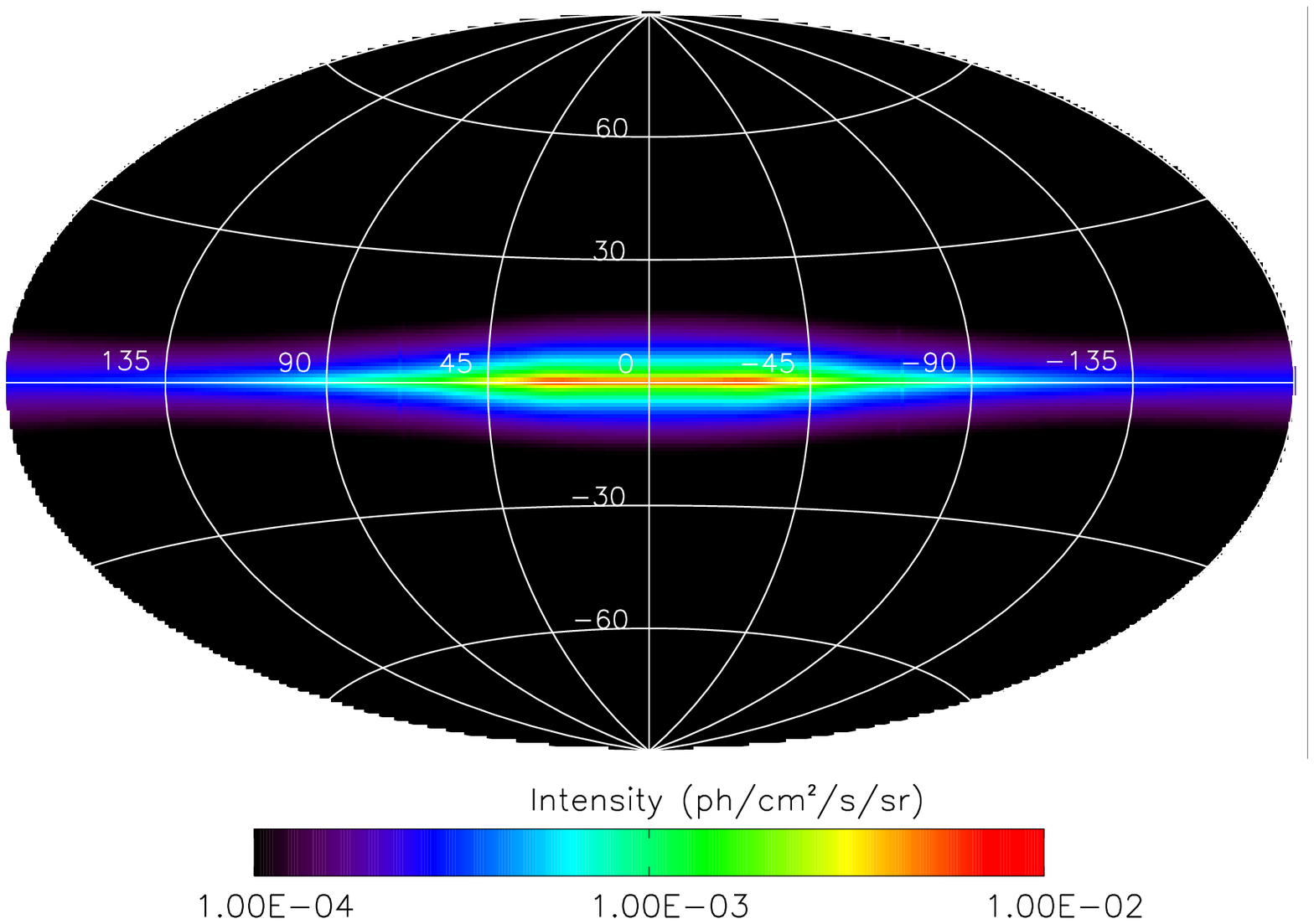}
\includegraphics[width=\columnwidth]{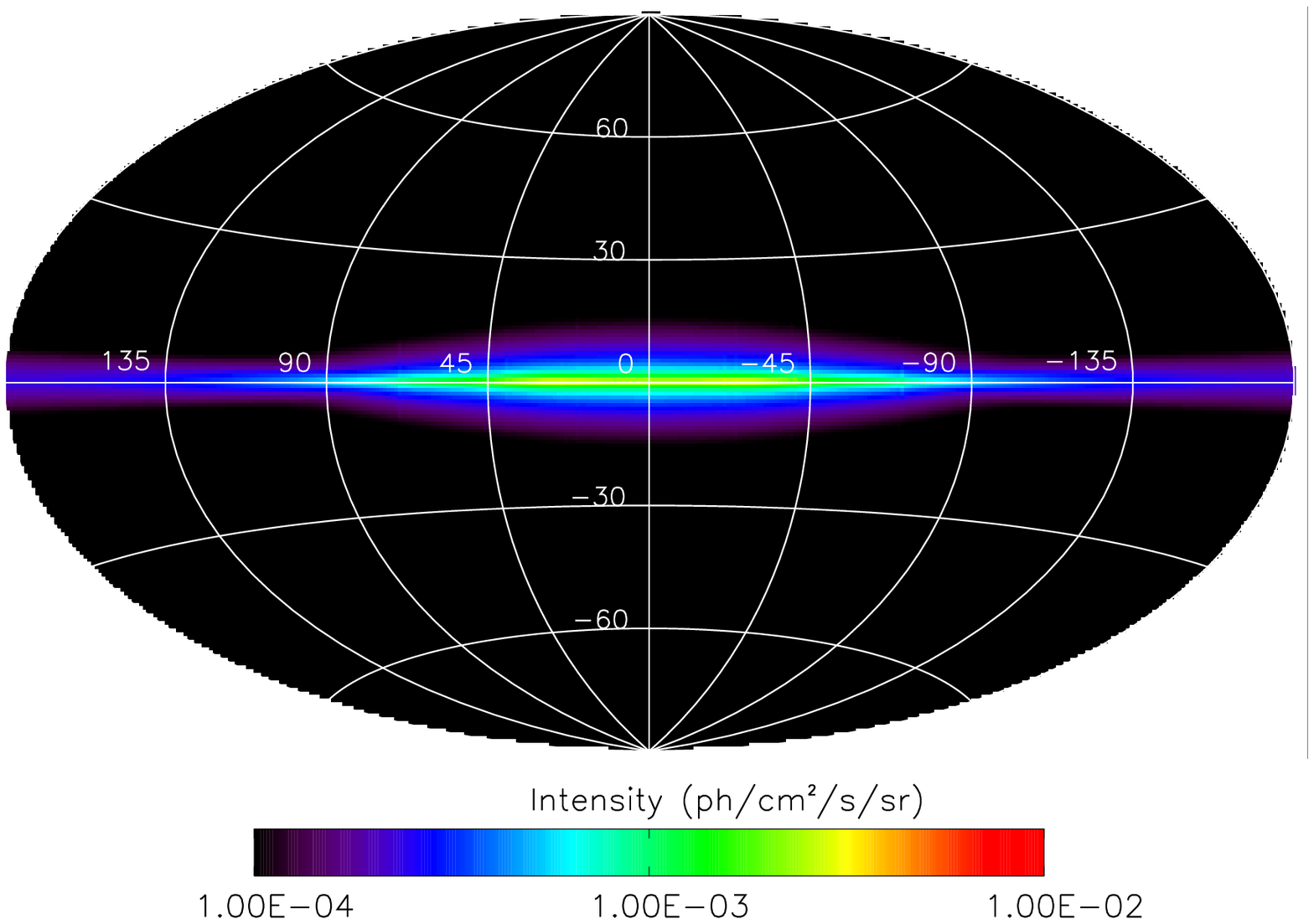}
\includegraphics[width=\columnwidth]{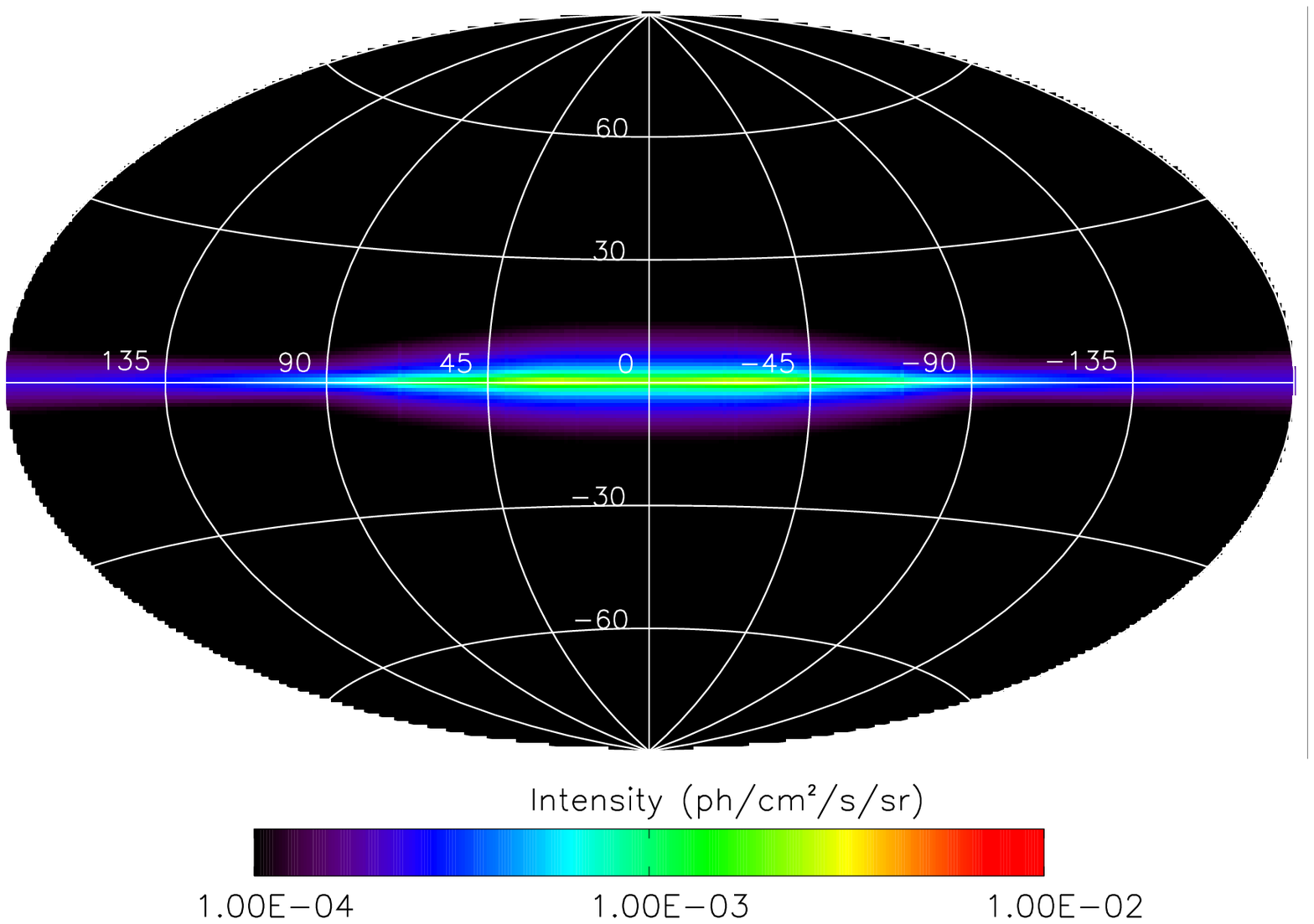}
\caption{Predicted 511\,keV intensity distributions for the annihilation of $^{44}$Ti positrons in each transport configuration. From top to bottom, the transport was assumed to be collisionless everywhere (case A, small diffusion coefficient), collisionless in the bulge and ballistic out of it (case B, inhomogeneous case), and ballistic everywhere (case C, large diffusion coefficient). The given intensities correspond to parapositronium annihilation only and were normalised to a positron injection rate of 10$^{43}$ e$^{+}$/s, with a positronium fraction of 0.95.}
\label{fig_44Ti_skymap}
\end{center}
\newpage
\end{figure}
\indent In its publicly-available version, GALPROP does not include annihilation processes. As explained in Sect. \ref{phys_anni}, positrons can annihilate with electrons through a variety of processes, the relative importance of which depend on the characteristics of the medium where the annihilation takes place. Yet, GALPROP does not include a fine description of the ISM at the typical scales of the various gas phases, but only the averaged and axisymetric gas distributions described above. At a given position in the space grid, the annihilation of positrons could not occur in a definite, well-identified phase but in an average medium resulting from the superposition of the large-scale distributions of molecular, atomic and ionised gas.\\
\indent We implemented positron annihilation in the following way: the rate of positron annihilation at a given position in the Galaxy is based on the rate of positrons being slowed down below $\sim$100\,eV, that is the energy below which positrons may experience charge exchange in-flight. Theoretically, this applies only to the neutral phases because in ionised phases positrons form Ps by radiative recombination or annihilate directly after complete thermalisation. In the warm ionised phase, the thermalisation from 100\,eV energies and the subsequent annihilation occur on short time scales compared to the slowing-down, so limiting the modelling to $\geq$100\,eV energies has no consequences on our results. In the hot phase, however, the thermal or near-thermal positrons have long lifetimes owing to the very low density. In the disk, these positrons are thus expected to be transported out of the hot cavities, very likely by strong turbulence, and to annihilate in the denser surrounding phases \citep{Jean:2006,Jean:2009}. This happens on time scales $\leq$1\,Myr, of the order of the slowing down time, and so once again limiting ourselves to $\geq$100\,eV energies will not strongly alter the results (especially since the cell size in our spatial grid is larger than the typical size of hot cavities, $\sim$100\,pc).\\
\indent The corresponding positron annihilation emissivity $q$ for position $\vec{r}$ and photon energy $E_{\gamma}$ reads
\begin{equation}
q(\vec{r},E_{\gamma})=\frac{s(E_{\gamma},f_{Ps})}{4\pi}\left[\varphi(\vec{r},E_{p}) \dot{E}_{p}(\vec{r},E_{p})\right]_{100\,\textrm{eV}}
\end{equation}
where $f_{Ps}$ is the Ps fraction, and the $s(E_{\gamma},f_{Ps})$ function gives the spectral distribution of the radiation. Yet, since we do not track the annihilation in individual ISM phases, we cannot directly model the annihilation spectrum over the Galaxy. We therefore fixed the spectral characteristics of the annihilation radiation: the Ps fraction was taken to be $f_{Ps}$=0.95, the 511\,keV line was assumed to have a Gaussian shape and a width of 6\,keV, and the Ps continuum was described by the \citet{Ore:1949} formula. We then focused on the intensity distribution over the sky, which is obtained by integrating the above emissivity along the line-of-sight for each direction to the sky\footnote{Note that our results can be rescaled to any other Ps fraction than the $f_{Ps}$=0.95 adopted: for instance, the 511\,keV intensities presented here can be converted to the case of a different Ps fraction $g_{Ps}$ after a multiplication by \mbox{(1-0.75$g_{Ps}$)/(1-0.75$f_{Ps}$).}}.\\
\indent We also implemented in our version of the GALPROP code the process of direct in-flight annihilation. In that case, it was implemented with its complete differential cross-section because it depends only on the bound and free electron density. However, we do not insist on that aspect because in-flight annihilation is negligible for MeV positrons.

\section{Simulation setups}
\label{simu}

\begin{figure}[!ht]
\vspace{1cm}
\begin{center}
\includegraphics[width=\columnwidth]{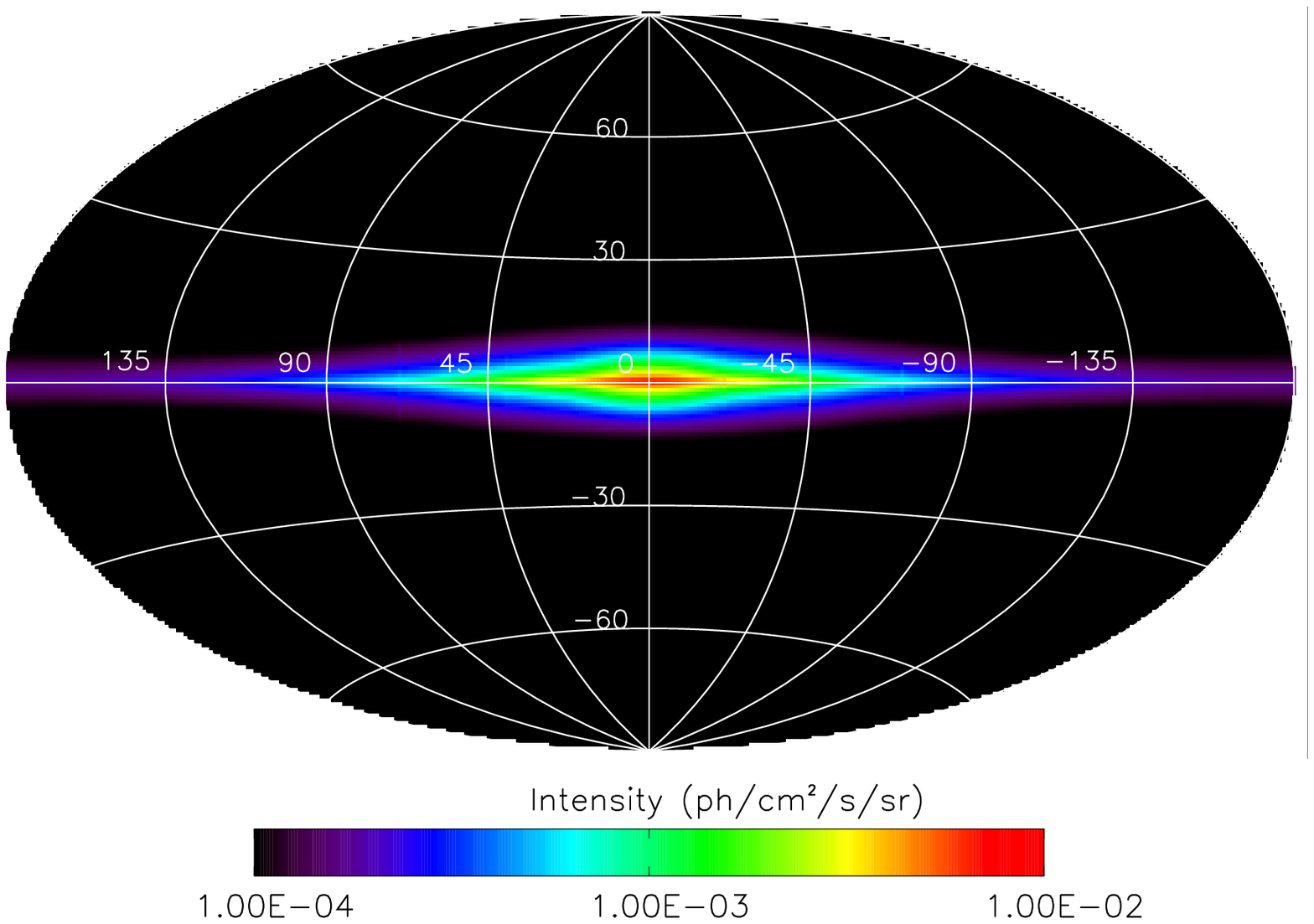}
\includegraphics[width=\columnwidth]{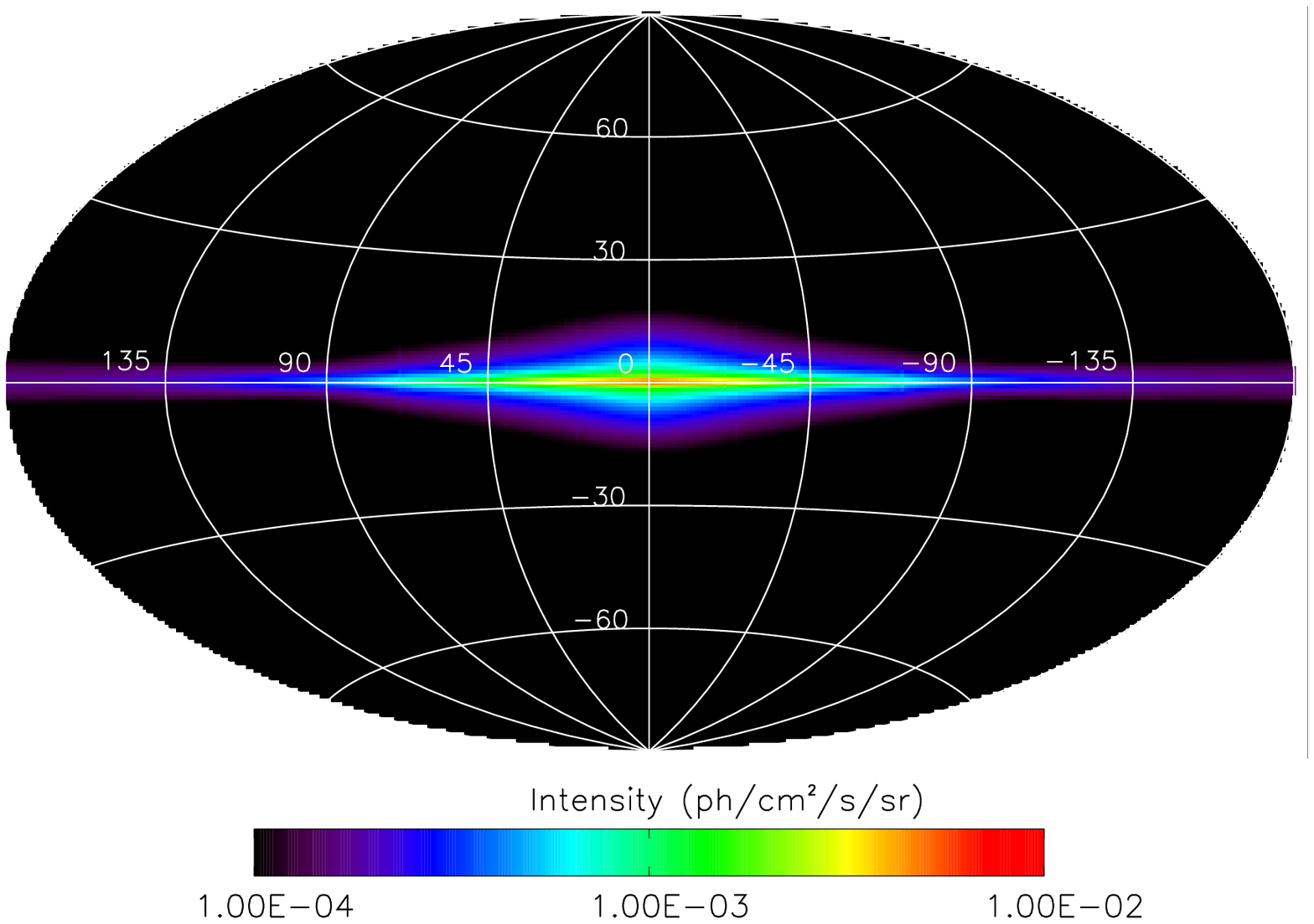}
\includegraphics[width=\columnwidth]{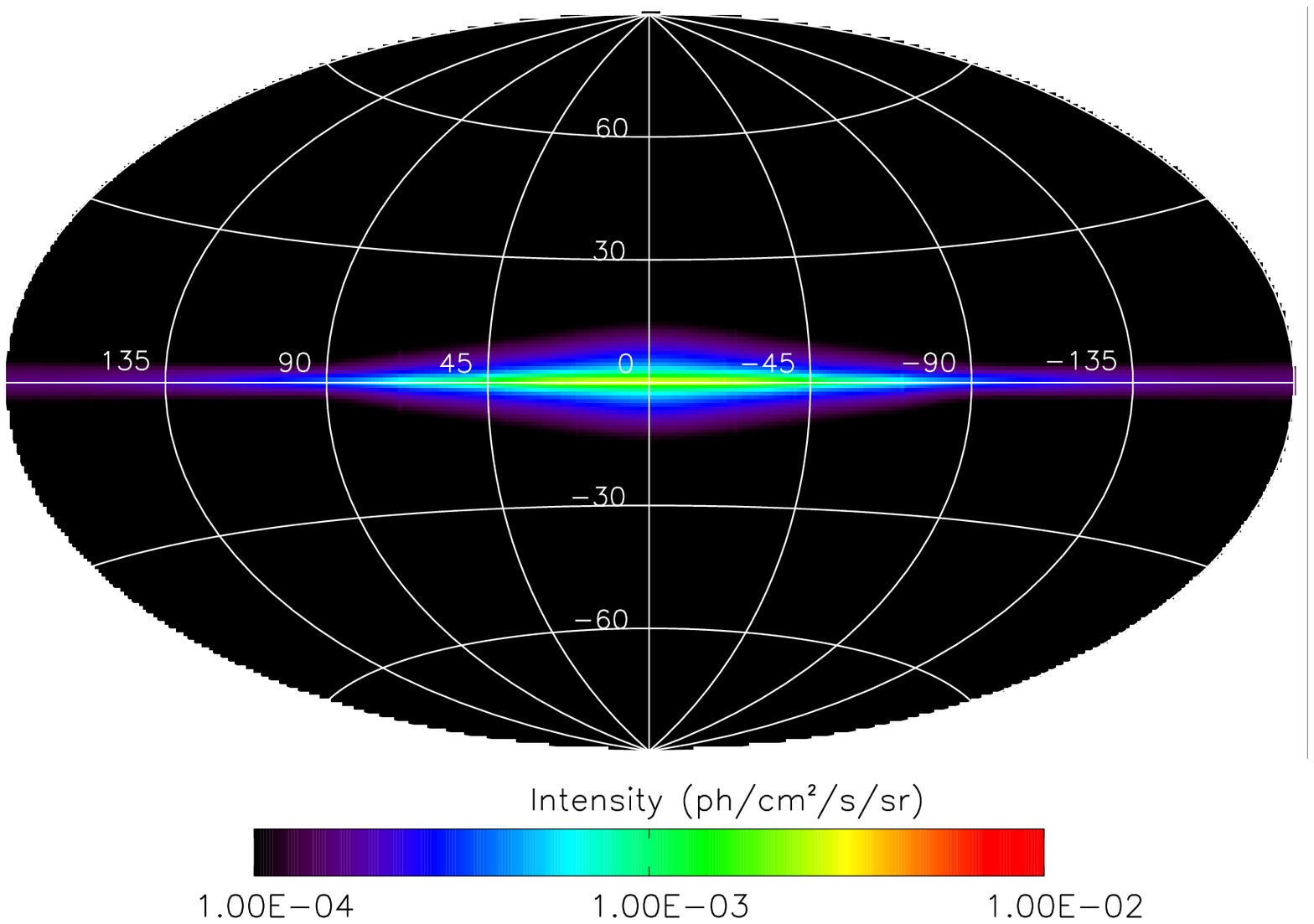}
\caption{Predicted 511\,keV intensity distributions for the annihilation of $^{56}$Ni positrons in each transport configuration. From top to bottom, the transport was assumed to be collisionless everywhere (case A, small diffusion coefficient), collisionless in the bulge and ballistic out of it (case B, inhomogeneous case), and ballistic everywhere (case C, large diffusion coefficient). The given intensities correspond to parapositronium annihilation only and were normalised to a positron injection rate of 10$^{43}$ e$^{+}$/s, with a positronium fraction of 0.95.}
\label{fig_56Ni_skymap}
\end{center}
\newpage
\end{figure}
\indent The numerical model described above allowed us to simulate the propagation of nucleosynthesis positrons and compute their annihilation emission for the entire Galactic system in a clear and consistent way. The predicted annihilation emissions basically rely on two inputs: the source spatial distribution and the propagation parameters. The source distributions were considered to be more firmly established than the transport conditions, so we fixed the source spatial parameters to the values given in Sect. \ref{src} and \ref{simu_src} and explored how the annihilation emission vary upon different prescriptions for the transport.\\
\indent All simulations presented in this paper were made in 2D cylindrical geometry. The spatial grid extends from 0 to 20 kpc in Galactocentric radius $r$, with a step size of 250\,pc, and from -4 to 4\,kpc in Galactic height $z$, with a step size of 100\,pc. This implies that the Galactic halo extends up to 4\,kpc on either side on the Galactic plane. The spatial boundary conditions at the edges of the spatial domain assume free particle escape, which means that the ultimate fate of positrons is either to annihilate in the disk or halo if they are slowed down below 100\,eV, or to escape if they diffuse far enough. The resolution of our space grid corresponds to $\sim$1-2\deg\ at the distance of the Galactic centre, which is about the size of the strongest peak in the 511\,keV signal. The energy grid runs from 100\,eV to 2\,MeV on a logarithmic scale, with 156 energy bins.

\subsection{Sources parameters}
\label{simu_src}

\indent In this section, we present the 2D source spatial distributions that were adopted for the different positron source. We saw in Sect. \ref{src} that the injection sites of radioactivity positrons follow a spatial distribution in the Galaxy that has three components: a star-forming disk consisting of an annulus and spiral arms (for parent isotopes coming from massive stars, their ccSNe, and also from prompt SNe Ia), together with an exponential disk with central hole and an ellipsoidal bulge containing most of the stellar mass (for parent isotopes coming from delayed SNe Ia).\\
\indent For the star-forming disk, we used the azimuthally-averaged radial profile of the Galactic star formation rate determined by \citet{Boissier:1999}; since massive stars have short lifetimes of a few Myr, their distribution and that of ccSNe closely follows that of the star forming activity. The vertical profile was assumed to be exponential with a scale height of 200\,pc.\\
\indent The stellar mass disk and bulge components were modelled by the functions:
\begin{align}
&f_{D}(r,z) = n_{D,0} \times \left( e^{-\frac{r}{R_{D,h}}}-e^{-\frac{r}{R_{D,c}}} \right) \times e^{-\frac{|z|}{Z_{D,h}}}\\
&f_{B}(r < R_{B,c},z) =  n_{B,0} \times e^{-\frac{1}{2} \left(\frac{r^2}{R_{B,h}^2} + \frac{z^2}{Z_{B,h}^2} \right) }\\
&f_{B}(r \geq R_{B,c},z) = n_{B,0} \times e^{-\frac{1}{2} \left(\frac{r^2}{R_{B,h}^2} + \frac{z^2}{Z_{B,h}^2} \right) } \times e^{-\frac{1}{2} (r-R_{B,c})^2}
\end{align}
where $(r,z)$ are the Galactocentric distance and the height above the Galactic plane and the indices $D/B$ stand for disk and bulge, respectively. These formula approximate the usual prescriptions for the stellar disk and bulge shapes \citep[see for instance][]{Robin:2003}. Each component $D/B$ is described by 1 density normalisation factor $n_{0}$ and 3 geometrical parameters $(R_{h},Z_{h},R_{c})$: a scale radius and height, a cutoff radius. We adopted the following triplets: (2.5,0.3,1.3) for the disk, and (1.6,0.4,2.5) for the bulge, in units of kpc. The normalisation factors are computed from the data given in Sect. \ref{src} for each parent radio-isotope.

\subsection{Diffusion parameters}
\label{simu_diff}

\indent The recent theoretical studies reviewed in Sect. \ref{phys_trans} could not conclusively determine if the transport of low-energy positrons is ballistic or collisionless. Strong arguments were presented in favour of a ballistic transport, in particular in the neutral phases, but these conclusions are not yet backed by solid experimental evidence. We therefore tested two extreme transport configurations in the diffusion approximation, with small and large coefficients corresponding to collisionless and ballistic transport, respectively. This allowed us to assess the impact on the resulting annihilation emission. In addition, we simulated an intermediate case of inhomogeneous diffusion, where the transport was assumed to be collisionless in the Galactic bulge and ballistic in the Galactic disk. Below, we present these transport scenarios in more detail, including the diffusion coefficients used in the code. We recall here that GALPROP does not include a fine description of the ISM with distinct phases, and so diffusion should be implemented with average properties over a few 100\,pc scales. In addition, we assumed in all cases an isotropic diffusion, but anisotropies may well exist for diffusion along the Galactic plane and off that plane towards the halo, or along spiral arms and across them.\\
\indent Case A: We tested the scenario of a collisionless transport with homogeneous properties over the whole Galaxy. For the ionised phases of the ISM, this option actually remains partly open in the work of \citetalias{Jean:2009} and is invoked in \citetalias{Higdon:2009} (see Sect. \ref{phys_trans}). As a limiting case, we assumed that it holds for all ISM phases over the entire Galaxy. In this scenario, we implicitly assumed the existence everywhere in the Galaxy of MHD turbulence with a sufficient energy density in the required wavelength range, so that low-energy positrons can be efficiently scattered. The corresponding random walk process was then treated in the diffusion approximation with a coefficient of the form:
\begin{align}
\label{eq_diffwave}
&D_{w}(R) = \beta D_{0} \left( \frac{R}{R_{0}} \right)^{\delta} \\
&D_{w}(R) = \,3.7 \times 10^{27} \, \textrm{cm}^{2}\,\textrm{s}^{-1} \times \beta \left( \frac{R}{1\,\textrm{MV}} \right)^{0.33}
\end{align}
where $R=p/e$ is the rigidity of the particle and $\beta=v/c$. The adopted normalisation factor $D_{0}$ as well as the spectral index $\delta$ are taken from cosmic-ray propagation studies interpreting local cosmic-ray measurements at $\sim$1-100\,GeV in the frame of a Galactic cosmic-ray propagation model. The spectral dependence of the diffusion coefficient is assumed to hold down to $\sim$MeV energies, which so far remains unproven (see Sect. \ref{phys_trans}). On the other hand, the resulting diffusion coefficient is small enough that the typical range of a MeV positron in a 1\,cm$^{-3}$ density medium is of order $\sim$10\,pc, meaning that nucleosynthesis positrons annihilate close to their sources (technically, they hardly escape the cell where they were injected). We assumed that diffusion is homogeneous over the Galactic volume, which is equivalent to assuming that the characteristics of magnetic turbulence are homogeneous over the Galaxy.\\
\indent Case B: We then tested an intermediate, inhomogeneous transport scenario, where the transport is collisionless in the Galactic bulge and ballistic elsewhere, notably in the Galactic disk. The limit between the two regions is set at a Galactocentric radius of about 3\,kpc. The form of the diffusion coefficient in the ballistic case is given below. This scenario actually is motivated by the observation that the bulge and the disk have different ISM compositions: the bulge is largely dominated by the hot ionised phase, while the disk is mostly filled with neutral gas. Estimates for the filling factors of the warm neutral medium, warm ionised medium, and hot medium are respectively (0.2,0.1,0.7) for the bulge and (0.5,0.3,0.2) for the disk, while the molecular medium and cold neutral medium occupy comparatively negligible volumes \citep{Jean:2006,Higdon:2009}. From Sect. \ref{phys_trans}, this difference may imply that positrons would be more efficiently confined in the bulge by small-scale MHD waves in the predominant ionised phases, while they would stream more easily in the disk where damping mechanisms prevent small-scale MHD perturbations in the predominant neutral phases. In reality, large-scale diffusion properties over a few 100\,pc scales reflect the actual ISM composition in each Galactic region (like the fact that the ionised phases of the disk occupy 50\% of the volume and thus would moderate the easier streaming experienced in the neutral phases). In our simulations, however, we pushed the contrast between bulge and disk to the maximum and set the former to be fully collisionless and the latter to be fully ballistic. This choice clearly is biased towards reproducing the high observed bulge-to-disk 511\,keV luminosity ratio, but we will see that even with that assumption, our predictions do not agree with the INTEGRAL/SPI results.\\
\indent Case C: Last, we tested the scenario of a ballistic transport with homogenous properties over the whole Galaxy. This case is suggested by the work of \citetalias{Jean:2009}, who concluded that $\sim$MeV positrons are likely not scattered by MHD turbulence whatever the ISM phase. In this situation, $\sim$MeV positrons have large mean free paths of several kpc in terms of pitch angle scattering. On small scales, the transport is anisotropic and positrons follow magnetic field lines in a ballistic motion. On larger scales, however, we expect an isotropization of the transport by strong random fluctuations of the Galactic magnetic field on typical scales $\sim$100\,pc \citep[an argument that is commonly invoked in cosmic-ray studies; see for instance][]{Ptuskin:2006}. In the ballistic scenario, we therefore approximate the transport by a diffusion with a characteristic mean free path $L_{B}$, the largest scale of magnetic fluctuations. The coefficient then takes the form
\begin{align}
\label{eq_diffcoll}
&D_{c}(R) = \frac{1}{3} \beta c L_{B} \\
&D_{c}(R) = \,3.1 \times 10^{30} \, \textrm{cm}^{2}\,\textrm{s}^{-1} \times \beta \left( \frac{L_{B}}{100\,\textrm{pc}} \right) 
\end{align}
In this case, the diffusion coefficient has a very modest rigidity/energy dependence down to a few 10\,keV, and the range of particles is therefore imposed by the energy dependence of the slowing-down processes. We assumed that diffusion is homogeneous over the Galactic volume, which is equivalent to assuming that large-scale magnetic turbulence is homogeneous over the Galaxy.

\section{The results}
\label{results}

\indent In this section, we present the predicted Galactic annihilation emission of $^{26}$Al, $^{44}$Ti, and $^{56}$Ni positrons, for the various transport configurations considered. We compare these to one of the latest models of the allsky 511\,keV emission obtained from INTEGRAL/SPI observations. Last, we discuss our results in light of other recent theoretical studies.
\begin{figure}[!ht]
\vspace{1cm}
\begin{center}
\includegraphics[width=\columnwidth]{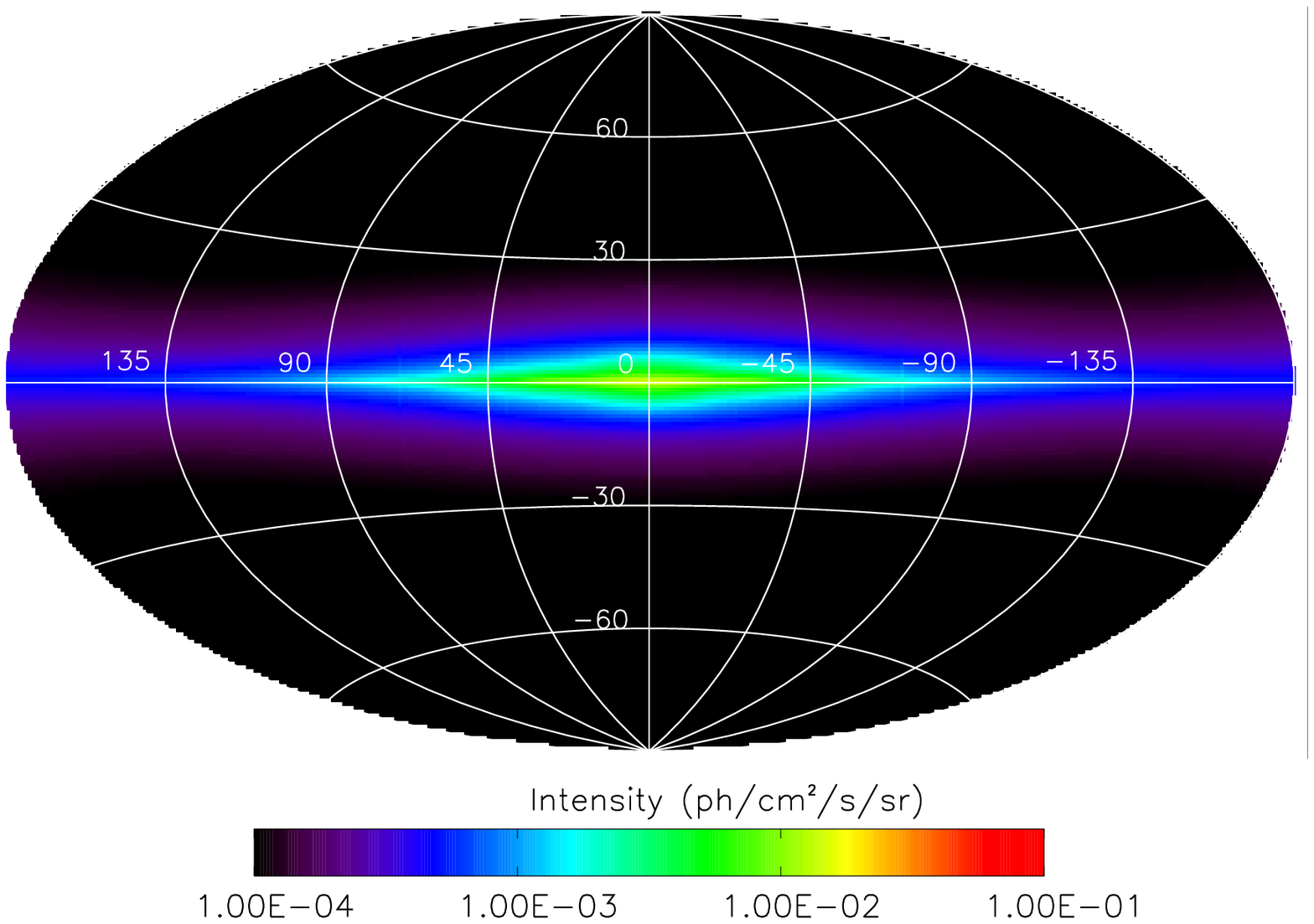}
\includegraphics[width=\columnwidth]{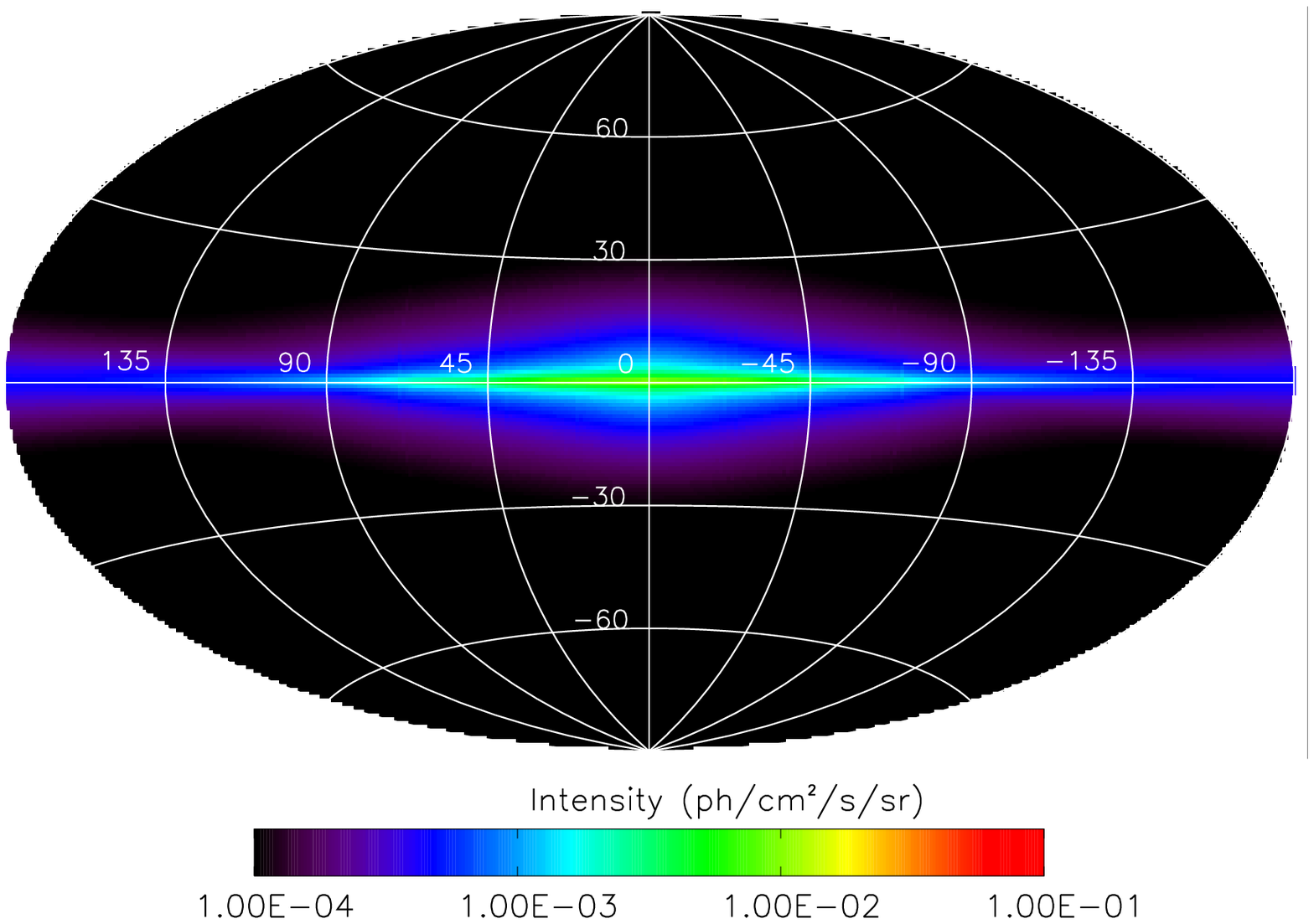}
\includegraphics[width=\columnwidth]{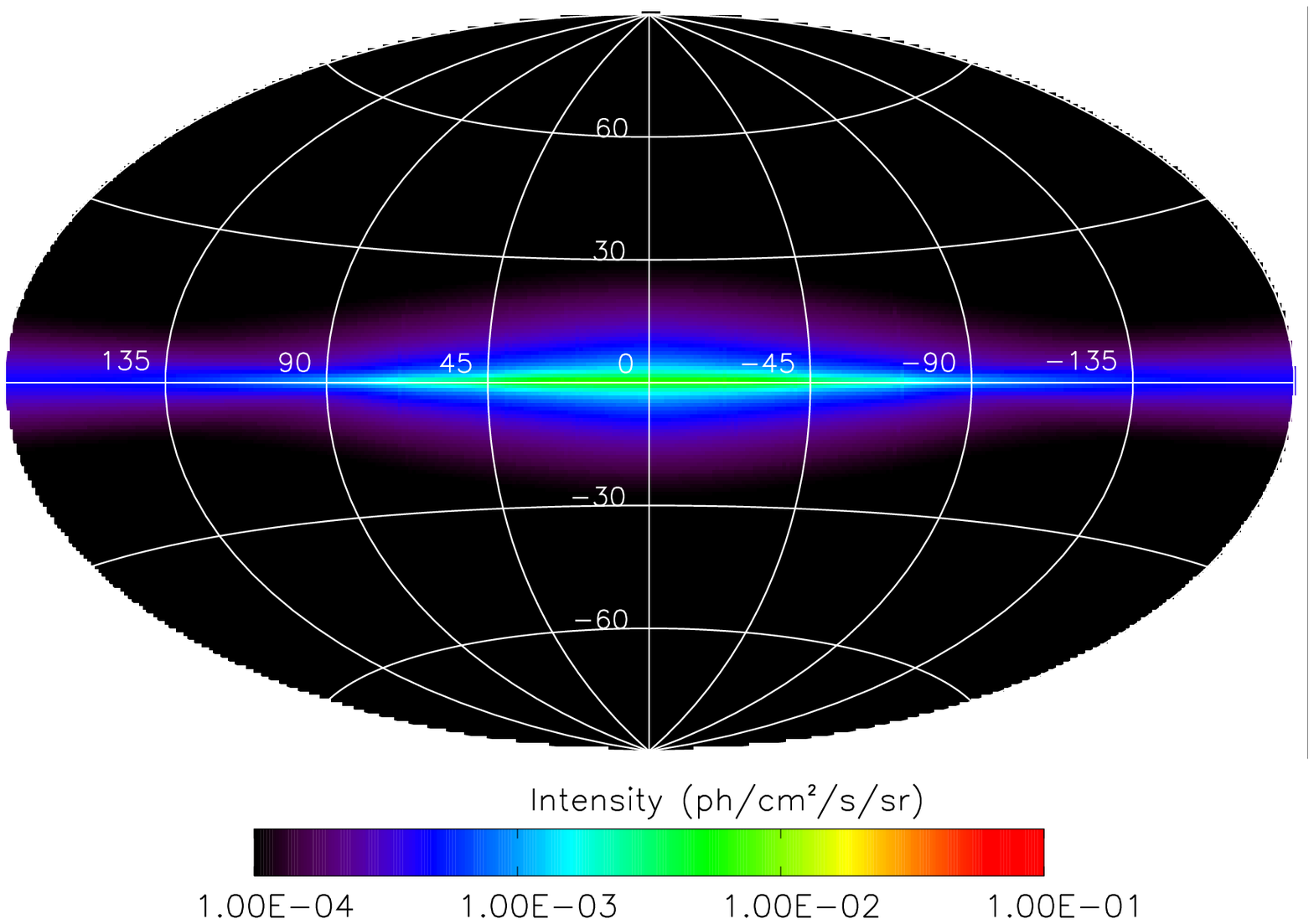}
\caption{Predicted 511\,keV intensity distributions for the annihilation of all nucleosynthesis positrons, using the estimated mean positron injection rates. From top to bottom, the transport was assumed to be collisionless everywhere (case A, small diffusion coefficient), collisionless in the bulge and ballistic out of it (case B, inhomogeneous case), and ballistic everywhere (case C, large diffusion coefficient). The given intensities correspond to the 511\,keV emission from parapositronium and direct annihilation, assuming a positronium fraction of 0.97.}
\label{fig_sum_skymap}
\end{center}
\newpage
\end{figure}

\subsection{Predicted intensity distributions}
\label{results_pred}

\begin{figure}[!ht]
\vspace{0.5cm}
\begin{center}
\includegraphics[width=\columnwidth]{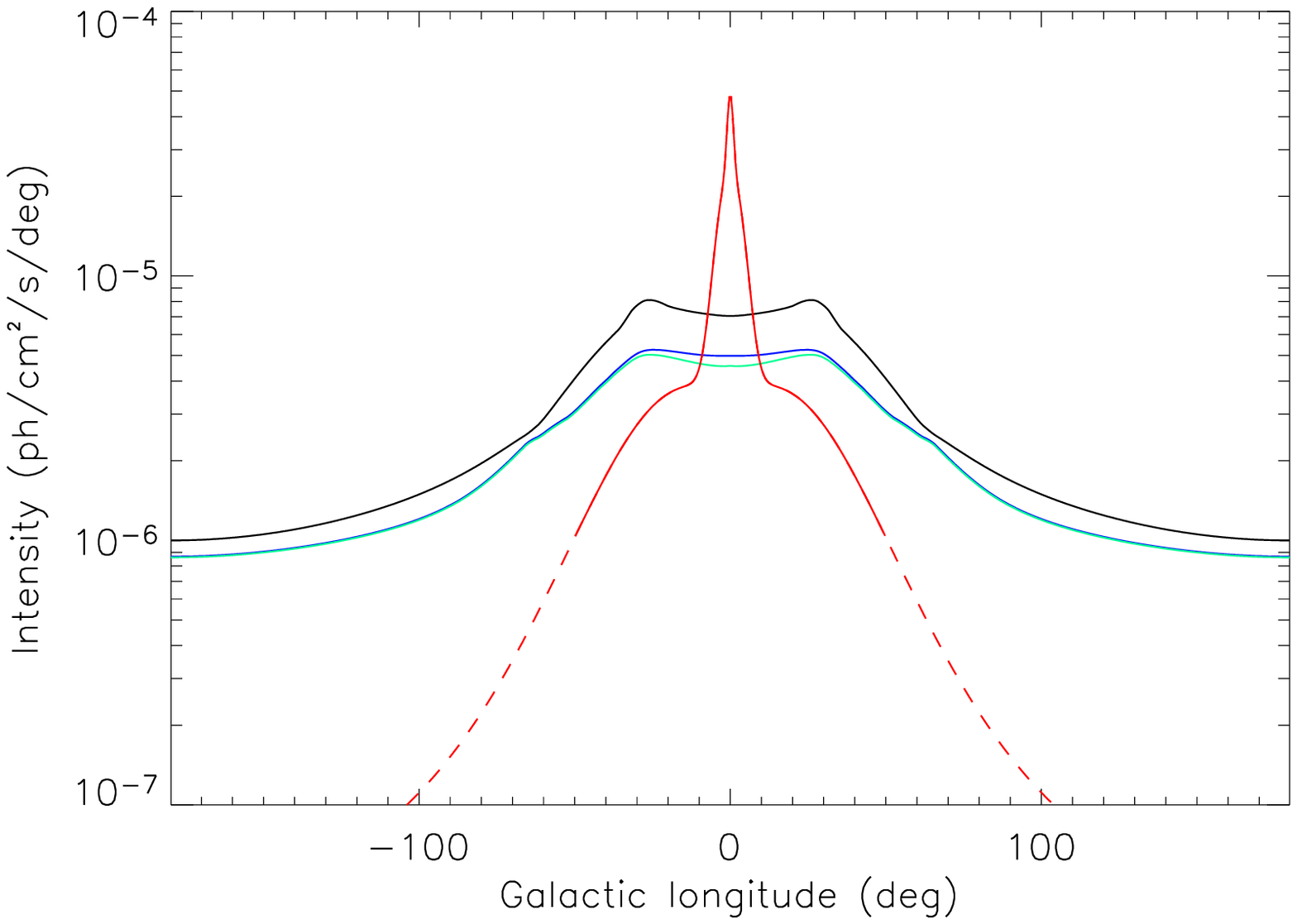}
\includegraphics[width=\columnwidth]{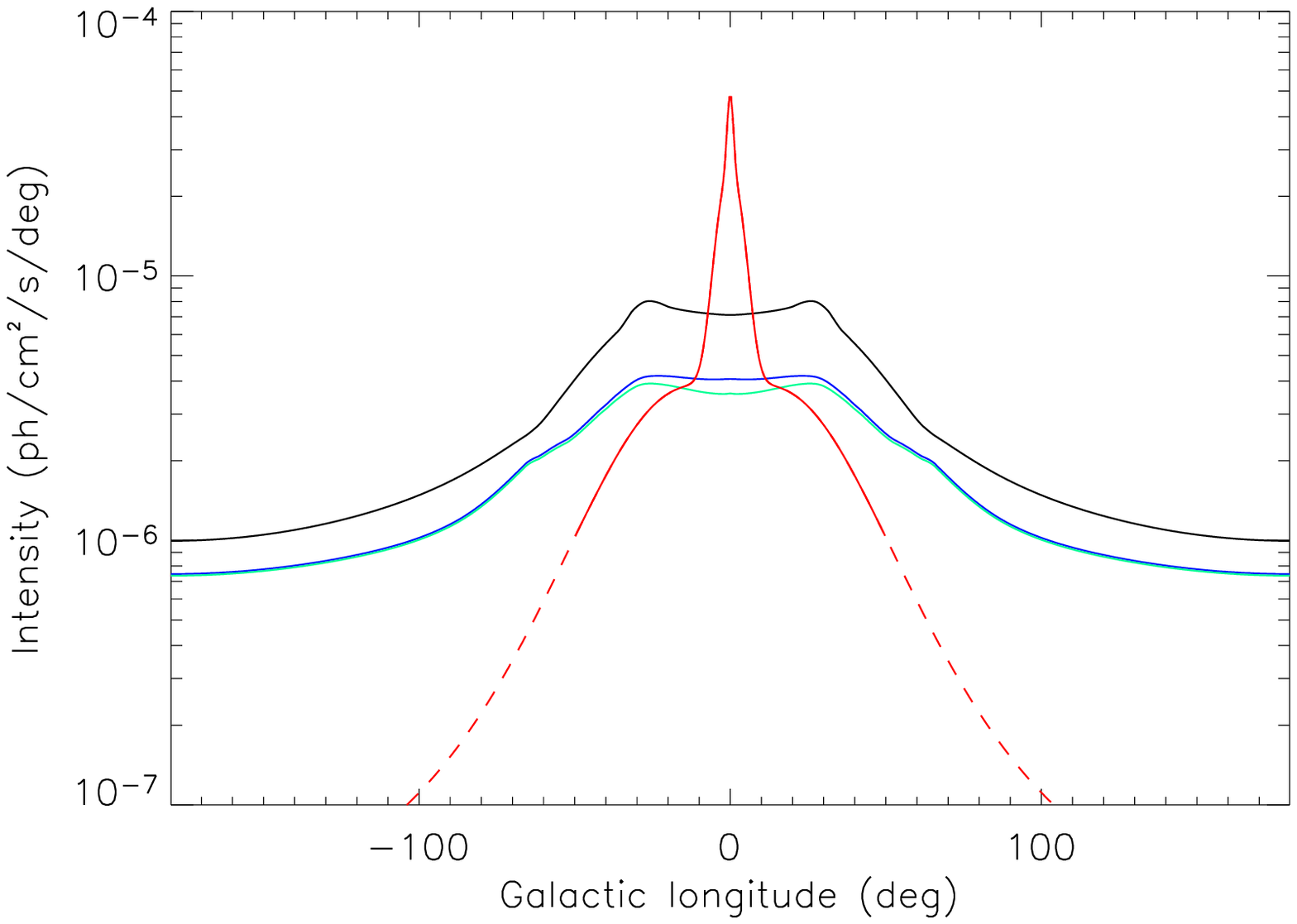}
\includegraphics[width=\columnwidth]{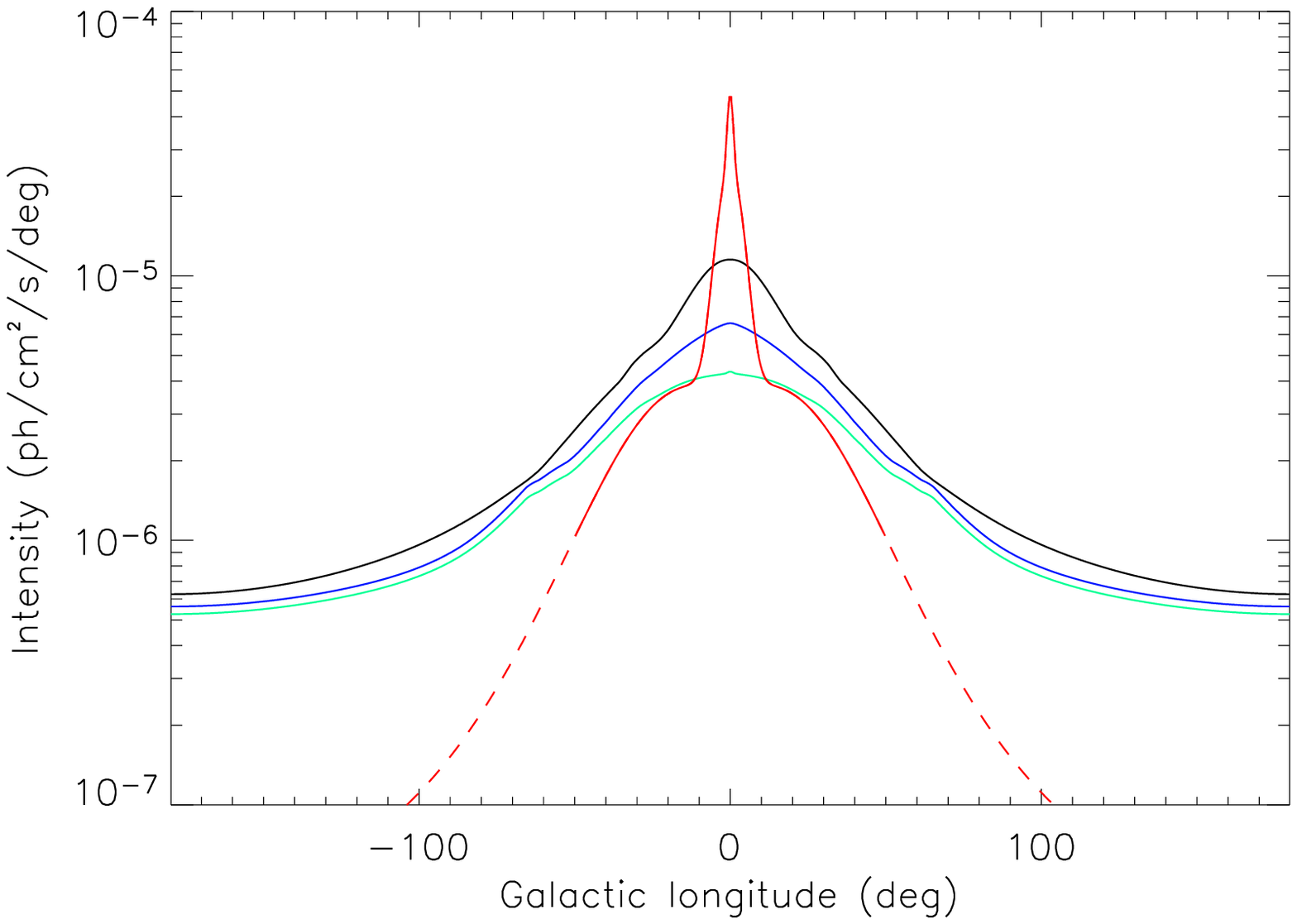}
\caption{Longitude profiles of the 511\,keV emission from $^{26}$Al, $^{44}$Ti, and $^{56}$Ni positrons (top, middle, and bottom panel respectively). Black, blue, and green curves correspond to transport configuration A, B, and C respectively. The red curves are the profile of a 511\,keV sky model obtained from model-fitting to the INTEGRAL/SPI observations (the dashed part representing the longitude range over which the intensity is poorly constrained). The given intensities correspond to parapositronium annihilation only and were normalised to a positron injection rate of 10$^{43}$ e$^{+}$/s, with a positronium fraction of 0.95. The profiles were integrated over a $\pi/9$\,rad latitude band centred on the Galactic plane.}
\label{fig_each_profile}
\end{center}
\end{figure}

\indent We present in Figs. \ref{fig_26Al_skymap}, \ref{fig_44Ti_skymap}, and \ref{fig_56Ni_skymap} the 511\,keV intensity distributions obtained in each transport configuration for $^{26}$Al, $^{44}$Ti, and $^{56}$Ni positrons, respectively. To facilitate the comparison, each skymap shows the parapositronium annihilation only, for a positronium fraction of 0.95 and a total Galactic positron injection rate of 10$^{43}$ e$^{+}$/s, of the same order as that inferred from observations. To emphasise the trends and differences, we show in Fig. \ref{fig_each_profile} the longitude profiles for each skymap. To help interpreting these results, we list in Table \ref{tab_lumi} the annihilation fraction and the annihilation luminosities for the bulge, disk, and entire Galaxy.\\
\indent For a same transport scenario, the skymaps look pretty similar: the intensity is highest in the inner Galaxy and progressively fades away with increasing longitude until it has dropped by about an order of magnitude in the outer Galaxy. The impact of the transport scenario on the resulting intensity distributions remains limited despite varying the diffusion coefficient by three orders of magnitude.\\
\indent In the fully collisionless scenario, positrons have very short ranges and they all annihilate close to their injection sites. The intensity distributions in that case reflect the source distributions (and are therefore strongly driven by our assumptions about the latter). The normalised intensity profiles for $^{44}$Ti and $^{26}$Al positrons are almost identical, because they have a similar source distribution dominated by the star-forming disk. They exhibit a plateau over the inner $\pm 30\deg$ in longitude that actually corresponds to the molecular ring at a Galactocentric radius of $\sim 4.5$\,kpc. In contrast, the emission from $^{56}$Ni positrons peaks at the Galactic centre since $^{56}$Ni positrons are preferentially released in the inner regions. It is interesting to note that when $^{56}$Ni positrons are confined to the vicinity of their sources (scenario A), they do not give rise to a highly-peaked 511\,keV signal from the inner bulge, resulting for instance from massive annihilation in the strong concentration of molecular gas in the central $\sim$200\,pc; instead, they seem to annihilate in a more distributed fashion over the entire bulge. The results of the fully collisionless scenario can be expected to be representative of what is obtained when positrons are injected into the ISM with a much smaller average energy than that of their original $\beta$-decay spectrum, whatever the transport conditions (this is especially relevant to $^{56}$Ni positrons that may be considerably slowed-down on their way out of the stellar ejecta). In that case, positrons lose their initially-small kinetic energy over short distances and thus annihilate close to their sources.\\
\indent As the diffusion efficiency is increased from scenario A to B and then C, the overall intensity goes down because an increasing fraction of the positrons escape the system. In the fully ballistic scenario, a fraction of 30-40\% of the positrons manage to stream out of the Galaxy. The fact that the annihilation fractions are lower for $^{44}$Ti and $^{56}$Ni positrons than for $^{26}$Al positrons is due to their higher average energy at injection, which allows them to survive longer and thus reach the boundaries of the system. This can also be seen in the longitude profiles, where the intensity drop between scenario A and B/C is stronger for $^{44}$Ti than for $^{26}$Al. The longitude profiles are not just shifted down as the diffusion efficiency is increased; the drop actually is stronger for the inner regions than for the outer Galaxy. This can be explained as follows: a larger diffusion coefficient allows more positrons to escape perpendicularly to the plane towards the halo, which lowers the overall luminosity; in the same time, the most energetic positrons spread out radially from the innermost regions, which flattens the longitude profile.\\
\indent For $^{44}$Ti and $^{26}$Al positrons, there is almost no difference between scenarios B and C. This is because there is almost no bulge contribution to the injection of these positrons, and so their emission is mainly influenced by the transport conditions in the disk (which are the same in scenarios B and C). In both cases, the intensity plateau remains clearly apparent and this shows the predominant role of the molecular ring, where positrons can slow down and annihilate very efficiently in the dense gas. For $^{56}$Ni positrons, there is a clear change of the emission profile from scenario A to B and then C. In scenario B, it is interesting to note that despite a low diffusion coefficient in the inner regions, half of the bulge positrons manage to escape (see Table \ref{tab_lumi}). A fraction of these are transported out to the disk and especially the molecular ring. As we move to scenario C, bulge positrons can escape much more easily towards the halo and fewer of them will feed the disk, the luminosity of which slightly diminishes. The decrease of the disk luminosity between scenarios B and C for $^{56}$Ni positrons actually reflects the decrease of the molecular ring luminosity, caused by a reduction of the supply by the bulge.\\
\indent The case of $^{56}$Ni positrons reveals an important difference between the bulge and the disk in terms of transport. Moving from fully collisionless to fully ballistic transport (scenario A to C), the luminosity of the bulge drops by $\sim$70\% while the luminosity of the disk decreases only by $\sim$30\% (see Table \ref{tab_lumi}). It is obviously easier to escape the bulge than the disk. This is due to the relative size and emptiness of the bulge, and to the fact that the spatial distribution of $^{56}$Ni positron sources in the bulge has a relatively high scale height of 400\,pc. In contrast, the disk harbours a dense molecular ring that positrons cannot easily leave, especially since most of them are injected directly in the molecular ring through massive stars, ccSNe, and prompt SNe Ia. The molecular ring is in the same time a major source and a major trap for nucleosynthesis positrons.\\
\indent Based on these results for each source, we computed a set of full emission models by adding for each transport scenario the contributions of all three isotopes weighted by their respective mean positron injection rate presented in Sect. \ref{src}. To allow direct and meaningful comparison with the data, we rescaled the outputs of our transport code so that the resulting intensities correspond to all 511\,keV processes (annihilation through parapositronium and direct annihilation of thermalised positrons) for a positronium fraction of 0.97 \citep[the value inferred from the observations by][]{Jean:2006}. The eventual skymaps are shown in Fig. \ref{fig_sum_skymap}, and the corresponding longitude profiles in Fig. \ref{fig_sum_profile}. Due to the preponderance of $^{56}$Ni positrons, the intensity profile peaks at the position of the Galactic centre. Yet, the corresponding B/D luminosity remains low, with a value of order 0.2-0.4 depending on the transport configuration. This is an order of magnitude below the values of 2-6 inferred from the gamma-ray observations. In the following, we will compare our predictions to observations and show that nucleosynthesis positrons can account for only a fraction of the measured annihilation signal.

\subsection{Comparison with INTEGRAL/SPI data}
\label{results_integral}

\begin{figure}[!t]
\begin{center}
\includegraphics[width=\columnwidth]{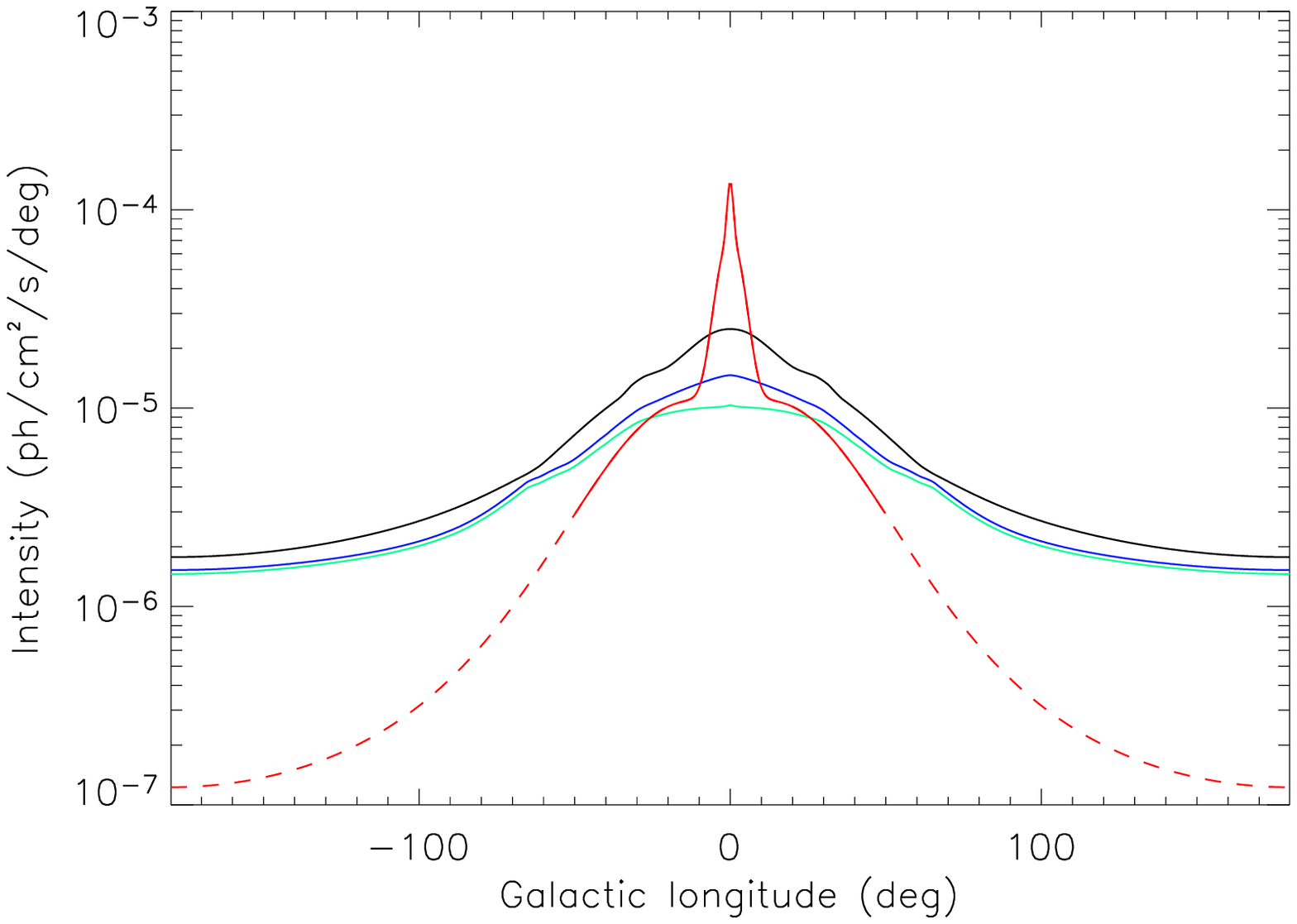}
\caption{Longitude profiles of the 511\,keV emission of all nucleosynthesis positrons for each transport configuration. The color coding is the same as in Fig. \ref{fig_each_profile}. The intensities correspond to the 511\,keV emission from parapositronium and direct annihilation, for a positronium fraction of 0.97.}
\label{fig_sum_profile}
\end{center}
\end{figure}

\begin{table}[t]
\begin{center}
\caption{Results of the fits of our predicted intensity distributions to about 7 years of INTEGRAL/SPI observations at 511\,keV.}
\begin{tabular}{|c|c|c|c|c|}
\hline
\celltspace Case & Inner & Outer & Disk & MLR \cellbspace \\
\hline 
\celltspace A & 1.34 $\pm$0.17 & 5.52 $\pm$0.35 & 17.1 $\pm$1.7 (0.61) & 2687.9 \\
\celltspace B & 1.31 $\pm$0.17 & 6.01 $\pm$0.32 & 18.8 $\pm$1.7 (0.88) & 2687.8 \\
\celltspace C & 1.29 $\pm$0.17 & 6.42 $\pm$0.30 & 19.3 $\pm$1.8 (1.03) & 2684.4 \\
\celltspace W & 1.44 $\pm$0.17 & 4.57 $\pm$0.33 & 15.6 $\pm$1.4 & 2692.3 \cellbspace \\
\hline
\end{tabular}
\label{tab_fit}
\end{center}
Note to the table: Cases A,B,C correspond to our models of all nucleosynthesis positrons with their mean injection rates and the three transport scenario considered. Case W correspond to the best-fit reoptimised Weidenspointner model. Columns 2-4 gives the fitted fluxes in $10^{-4}$\,ph\,cm$^{-2}$\,s$^{-1}$ for the inner bulge, outer bulge, and disk components, respectively. The numbers in parentheses are the scaling factors applied to our models in the fit. Column 6 gives the Maximum Likelihood Ratio.
\end{table}
\indent From about 7 years of INTEGRAL/SPI data, we repeated the analysis done in \citet{Weidenspointner:2008a} to determine by model-fitting the 511\,keV intensity distribution that best accounts for the gamma-ray observations. We used the same components adopted by these authors for their symmetric model: a superposition of two spheroidal Gaussian distributions for the bulge emission, and an exponential disk with central hole for the longitudinally-extended emission \citep[the young stellar disk model of][]{Robin:2003}. All components are assumed to be centred at the Galactic centre. We reoptimised iteratively some parameters of the model: first the widths of the Gaussians and then the disk scale length and height. The intensity distribution of our best-fit model is shown in Fig. \ref{fig_obs_skymap}, and it is similar to the results presented in \citet{Weidenspointner:2008a}. In agreement with other studies, the intensity is strongly concentrated in the inner Galaxy with a major bulge contribution. Such a model seems to suggest that there is little contribution from the disk at $|l| > 50\deg$, but this actually comes from an instrumental limitation. Most INTEGRAL/SPI exposure is on the inner Galaxy, which is also where the surface brightness of the 511\,keV signal is highest. Consequently, any best-fit model can be expected to provide a more reliable account of that part of the sky compared to higher longitude and/or latitude regions. Further away from the inner regions, the intensity predicted by the model is mainly driven by the assumed functional form for the fitted disk component. In other words, the intensity distribution at $|l| > 40-50\deg$ should be considered as presently unconstrained \citep[see Fig. 6 in][no significant 511\,keV emission is detected in $5\deg \times 5\deg$ sky pixels beyond $|l| > 40\deg$]{Bouchet:2010}. In the following, we will therefore focus on the inner $|l| < 50\deg$ when comparing our predicted intensity distributions to that derived from observations.\\
\indent Comparing this best-fit model to our predicted emission for all nucleosynthesis positrons, the intensity distributions appear to be very different. From Fig. \ref{fig_sum_profile}, even allowing for some renormalisation of the positron injection rates within the ranges of values presented in Sect. \ref{src}, it is obvious that the annihilation of nucleosynthesis positrons cannot account for the strong central peak observed in the inner $l= \pm 10\deg$. So at least in the frame of our model, an extra source of positrons is needed to explain the bulge emission. Outside the bulge, however, nucleosynthesis positrons can account satisfactorily for the emission in the $l= \pm 10-50\deg$ range, provided the positron injection rates are adjusted about the mean values we derived earlier (see below).\\
\indent To demonstrate this, we replaced the exponential disk by our predicted intensity distributions in a series of fits to the INTEGRAL/SPI data. The two bulge components of the best-fit model pictured in Fig. \ref{fig_obs_skymap} were used to account for the bulge emission (only their normalisation was fitted, the morphology parameters were not reoptimised). The fit results are presented in Table \ref{tab_fit}. Our models for the disk emission can account for the INTEGRAL/SPI data almost as well as the exponential disk of the Weidenspointner best-fit model (especially if one considers that the bulge was not optimised for our models). The difference in maximum likelihood ratio between our models A or B and the Weidenspointner best-fit model is about 4, which would be an insignificant difference even if only one parameter of the disk model had been modified (this also confirms the above statement that the high-longitude emission is poorly constrained, since our predicted intensity profiles clearly differ from that of the exponential disk model at $|l| > 50\deg$). On the contrary, using our disk models seem to require a higher outer bulge emission, by $\sim$40\% in the most extreme case (here again, it might be necessary to reoptimize the outer bulge width).\\
\indent In terms of information about positron injection and transport, our fits showed that the data cannot significantly distinguish between the different transport scenarios tested in that work, most likely because our predicted intensity distributions do not exhibit strong variations upon modifications of the transport conditions. The best models A and B were rescaled in the fit by 0.61 and 0.88 respectively. This could be interpreted as an indication that the positron injection rates were slightly overestimated. Applying the same downscaling to all positron sources would imply for instance escape fractions of 61-88\% and 3-4\% for the $^{44}$Ti and $^{56}$Ni positrons, respectively, which lie in the ranges of possible injection rates presented in Sect. \ref{src}. Yet, considering the number of uncertain parameters involved in the modelling of the Galactic transport and annihilation of positrons, plus the fact that the fit is strongly influenced by the as-yet unexplained contribution of the bulge, such a conclusion on the positron injection rate should not be taken as very solid.

\subsection{Comparison with other studies}
\label{results_compa}

\indent We now compare our results with those obtained by other authors. Most of the comparison will be done with the work of \citetalias{Higdon:2009}, since this is one of the most extensive studies on the subject of nucleosynthesis positrons.\\
\indent \citetalias{Higdon:2009} claim that differential propagation of nucleosynthesis positrons can explain all properties of the INTEGRAL/SPI observations. We reach a different conclusion, at least for the morphology of the annihilation emission. We identified several assumptions in the work of \citetalias{Higdon:2009} that may explain the discrepancy and review them in the following.\\
\indent \textit{$^{44}$Ti contribution}: They assumed that $^{44}$Ti positrons come mostly from SNe Ip, a peculiar subtype of SNe Ia. While the exact origin of the present-day solar abundance of the daughter nucleus $^{44}$Ca remains uncertain \citep{The:2006}, we want to stress that the observations available today favour a dominant contribution from ccSNe (higher yield and higher occurrence rate, see Sect. \ref{src_44ti}), hence an injection in the star-forming disk rather than in the central regions.\\
\indent \textit{SNe Ia distribution}: They used a spatial distribution of SNe Ia based only on stellar mass and composed of an exponential disk and a bulge. We also used similar distributions (except our exponential disk has a central hole), but we had in addition a component based on star formation rate (see Sect. \ref{src_44ti}), hence more contribution from the star-forming disk. This is somewhat alleviated by the fact that our stellar bulge and disk contribute equal rates of SNe Ia, while in their model the bulge is responsible for only $\sim$1/3 of all SNe Ia.\\
\indent \textit{Positron energies}: It is not clear whether the authors used the complete $\beta$-decay spectra or worked with typical energies such as 1\,MeV or the average energies of the $\beta$-decay spectra. This may have important consequences since most positrons actually have initial energies of a few 100\,keV only, which limits their range. In our study, the full $\beta$-decay spectra were used.\\
\indent \textit{Bulge transport conditions}: The authors argued that the majority of positrons created in the inner and middle bulge within R $\leq$ 1.5\,kpc are confined to that region because their scattering mean free path in the outer bulge is smaller and cause them to be reflected back. In addition, a significant fraction of the positrons created in the outer bulge would be preferentially transferred down to the inner and middle bulge. All these positrons trapped within R $\leq$ 1.5\,kpc would eventually diffuse into the cloudy labyrinths of the tilted disk and central molecular zone, and annihilate there. This was however not proven by a complete modelling of the entire region.\\
\indent Overall, there are many assumptions that favour a strong concentration of the positron creation and annihilation in the inner Galactic regions. On the other hand, our approach has its own restrictions, and we identified below the most important of them.\\
\indent The main limitation of our global diffusion model is its large-scale definition, with gas densities and diffusion properties being average values over several 100\,pc at least. In that frame, it is not possible to simulate inhomogeneous diffusion in a realistic, finely-structured ISM, that is to say to follow the propagation of positrons through successive, identified phases with different transport properties. This is something Monte-Carlo simulations could do from statistical descriptions of the ISM properties (Alexis et al. 2012, in prep.). Regarding simply the diffusion, it may be that the details of the transport through various ISM phases of $\sim$1-10\,pc sizes can be captured in large-scale diffusion coefficients valid over a few 100\,pc scales and being possibly different over large Galactic regions (like the bulge and the disk in our transport configuration B). Yet, in terms of energy losses, it might be that the average gas densities over a few 100\,pc scales are not representative of that experienced by positrons if the diffusion really is inhomogeneous over the various ISM phases and positrons sample some phases more than others. This might be quite important especially in the inner bulge, where most annihilation seems to occur. On the other hand, the sources of nucleosynthesis positrons are very likely distributed over kpc scales, even within the bulge. It seems improbable that diffusive transport starting from such a widespread injection could result in a highly-localized annihilation because of peculiar conditions in a few 100\,pc size region.\\
\indent Another disregarded effect if the role of the large-scale magnetic field topology. Whatever the nature of the transport, ballistic or collisionless, the propagation of positrons is thought to be anisotropic on small scales, being more efficient along field lines than across them. We argued that the strong random fluctuations of the Galactic magnetic field over $\sim$100\,pc scales would isotropize the propagation over large scales (the same argument is invoked for cosmic-ray propagation studies). Yet, the Galactic magnetic field is not fully random over large scales and has a global structure made of a toroidal part in the plane (likely following a spiral pattern) and a poloidal part out of it (possibly X-shaped). How this global structure impact the transport of positrons remains uncertain and many scenarios can be imagined \citep[see for instance][]{Prantzos:2006}. We note, however, that most positrons do not travel very far from their injection sites. Even when they are allowed to do so (large diffusion coefficient, transport configuration C), about a half of the positrons annihilate in sites that closely follow the initial distribution of sources (see Table \ref{tab_lumi}).\\
\indent Beyond the limitations of any particular model, there are several reasons why large-scale propagation of nucleosynthesis positrons may not occur even if interstellar conditions and magnetic field topology seem favourable. Most positrons actually have initial energies of a few 100\,keV only, and not 1\,MeV as usually assumed in rough estimates, and the energy loss rate through ionisation/excitation and Coulomb interactions increase with decreasing energy below 1\,MeV. This would by itself restrict the range of positrons whatever their injection site. But in addition, all source distributions have a star-forming disk component, which means that many positrons are injected in the high-density molecular ring and annihilate there. From these arguments and the results of our modelling, it seems unlikely that only nucleosynthesis positrons be responsible for the observed annihilation emission. A corollary of the above statements is that the extra source needed to account for the bulge annihilation emission should very likely be concentrated in the inner regions, if the positrons have initial energies in the 100\,keV-1\,MeV range.

\section{Conclusion}
\label{conclu}

\indent The Galaxy hosts a low-energy positron population that is revealed through a clear signature of 511\,keV and continuum annihilation coming predominantly from the inner regions. A likely source of such particles is the decay of radioactive species produced by the nucleosynthetic activity of our Galaxy.  We assessed the contribution of $^{26}$Al, $^{56}$Ni and $^{44}$Ti positrons to the observed gamma-ray signal by simulating their transport and annihilation using a GALPROP-based model of high-energy particle diffusion in our Galaxy. Starting from source spatial profiles based on typical distributions of massive stars and supernovae, we explored how the annihilation intensity distributions vary upon different transport prescriptions ranging from low-diffusion, collisionless propagation to high-diffusion, ballistic propagation.\\
\indent In the frame of our model, these extreme transport scenarios at the scale of the Galaxy result in very similar intensity distributions with small underlying bulge-to-disk luminosity ratios of $\sim$0.2-0.4, which is an order of magnitude below the values inferred from observations. The intensity profiles are determined to a relatively high degree by the adopted source distributions. This indicates that most positrons do not travel very far from their injection sites, even when they are allowed to do so. We estimate that at least about half of the positrons annihilate in sites close to their sources. As propagation efficiency is enhanced from fully collisionless to fully ballistic scenarios, the fraction of positron escaping the system increases from 0 to 30-40\%. The corresponding luminosity drop is more pronounced for the bulge than for the disk, owing to a relatively tenuous medium over most of the bulge volume and a high scale height of the bulge sources. In contrast, the disk harbours a dense molecular ring that positrons cannot easily leave, especially since most of them are injected there through massive stars, ccSNe, and prompt SNe Ia. Allowing for specific transport conditions in the bulge does not result in a higher bulge-to-disk ratio for the annihilation emission.\\
\indent Comparing to the INTEGRAL/SPI observations of the 511\,keV emission, we conclude that the annihilation of nucleosynthesis positrons cannot account for the strong central peak observed in the inner $l= \pm 10\deg$. An extra source of positrons is needed to explain the bulge emission, and it is very likely concentrated to the innermost regions if positrons have initial energies in the 100\,keV-1\,MeV range. This contrasts with recent claims that nucleosynthesis positrons can explain all properties of the INTEGRAL/SPI measurements. Outside the bulge, however, nucleosynthesis positrons can account satisfactorily for the extended disk-like emission, as confirmed by a fit of our predicted intensity distributions to the data.

\begin{acknowledgement}
We thank the referee Vladimir Dogiel for his helpful comments and suggestions. The SPI project has been completed under the responsibility and leadership of CNES. We are grateful to ASI, CEA, CNES, DLR, ESA, INTA, NASA and OSTC for their support. Pierrick Martin acknowledges support from the European Community via contract ERC-StG-200911.
\end{acknowledgement}

\bibliographystyle{aa}
\bibliography{/Users/pierrickmartin/Documents/MyPapers/biblio/Pulsars,/Users/pierrickmartin/Documents/MyPapers/biblio/SMC,/Users/pierrickmartin/Documents/MyPapers/biblio/CosmicRaySources,/Users/pierrickmartin/Documents/MyPapers/biblio/CosmicRayTransport,/Users/pierrickmartin/Documents/MyPapers/biblio/GalaxyObservations,/Users/pierrickmartin/Documents/MyPapers/biblio/DataAnalysis,/Users/pierrickmartin/Documents/MyPapers/biblio/Fermi,/Users/pierrickmartin/Documents/MyPapers/biblio/Books,/Users/pierrickmartin/Documents/MyPapers/biblio/SNobservations,/Users/pierrickmartin/Documents/MyPapers/biblio/26Al&60Fe,/Users/pierrickmartin/Documents/MyPapers/biblio/Positron,/Users/pierrickmartin/Documents/MyPapers/biblio/44Ti,/Users/pierrickmartin/Documents/MyPapers/biblio/Cygnus&CygOB2,/Users/pierrickmartin/Documents/MyPapers/biblio/Physics}

\begin{thebibliography}{55}
\expandafter\ifx\csname natexlab\endcsname\relax\def\natexlab#1{#1}\fi

\bibitem[{{Beacom} \& {Y{\"u}ksel}(2006)}]{Beacom:2006}
{Beacom}, J.~F. \& {Y{\"u}ksel}, H. 2006, Physical Review Letters, 97, 071102

\bibitem[{{Boissier} \& {Prantzos}(1999)}]{Boissier:1999}
{Boissier}, S. \& {Prantzos}, N. 1999, \mnras, 307, 857

\bibitem[{{Borkowski} {et~al.}(2010){Borkowski}, {Reynolds}, {Green}, {Hwang},
  {Petre}, {Krishnamurthy}, \& {Willett}}]{Borkowski:2010}
{Borkowski}, K.~J., {Reynolds}, S.~P., {Green}, D.~A., {et~al.} 2010, \apjl,
  724, L161

\bibitem[{{Bouchet} {et~al.}(2010){Bouchet}, {Roques}, \&
  {Jourdain}}]{Bouchet:2010}
{Bouchet}, L., {Roques}, J.~P., \& {Jourdain}, E. 2010, \apj, 720, 1772

\bibitem[{{Bronfman} {et~al.}(1988){Bronfman}, {Cohen}, {Alvarez}, {May}, \&
  {Thaddeus}}]{Bronfman:1988}
{Bronfman}, L., {Cohen}, R.~S., {Alvarez}, H., {May}, J., \& {Thaddeus}, P.
  1988, \apj, 324, 248

\bibitem[{{Chan} \& {Lingenfelter}(1993)}]{Chan:1993}
{Chan}, K.-W. \& {Lingenfelter}, R.~E. 1993, \apj, 405, 614

\bibitem[{{Chernyshov} {et~al.}(2010){Chernyshov}, {Cheng}, {Dogiel}, {Ko}, \&
  {Ip}}]{Chernyshov:2010}
{Chernyshov}, D.~O., {Cheng}, K., {Dogiel}, V.~A., {Ko}, C., \& {Ip}, W. 2010,
  \mnras, 403, 817

\bibitem[{{Cho} \& {Lazarian}(2003)}]{Cho:2003}
{Cho}, J. \& {Lazarian}, A. 2003, \mnras, 345, 325

\bibitem[{{Cho} {et~al.}(2002){Cho}, {Lazarian}, \& {Vishniac}}]{Cho:2002}
{Cho}, J., {Lazarian}, A., \& {Vishniac}, E.~T. 2002, \apj, 564, 291

\bibitem[{{Churazov} {et~al.}(2005){Churazov}, {Sunyaev}, {Sazonov},
  {Revnivtsev}, \& {Varshalovich}}]{Churazov:2005}
{Churazov}, E., {Sunyaev}, R., {Sazonov}, S., {Revnivtsev}, M., \&
  {Varshalovich}, D. 2005, \mnras, 357, 1377

\bibitem[{{Cordes} \& {Lazio}(2002)}]{Cordes:2002}
{Cordes}, J.~M. \& {Lazio}, T.~J.~W. 2002, ArXiv Astrophysics e-prints

\bibitem[{{Cox} {et~al.}(1986){Cox}, {Kruegel}, \& {Mezger}}]{Cox:1986}
{Cox}, P., {Kruegel}, E., \& {Mezger}, P.~G. 1986, \aap, 155, 380

\bibitem[{{Dickey} \& {Lockman}(1990)}]{Dickey:1990}
{Dickey}, J.~M. \& {Lockman}, F.~J. 1990, \araa, 28, 215

\bibitem[{{Diehl} {et~al.}(2006){Diehl}, {Halloin}, {Kretschmer}, {Lichti},
  {Sch{\"o}nfelder}, {Strong}, {von Kienlin}, {Wang}, {Jean}, {Kn{\"o}dlseder},
  {Roques}, {Weidenspointner}, {Schanne}, {Hartmann}, {Winkler}, \&
  {Wunderer}}]{Diehl:2006}
{Diehl}, R., {Halloin}, H., {Kretschmer}, K., {et~al.} 2006, \nat, 439, 45

\bibitem[{{Ferri{\`e}re} {et~al.}(2007){Ferri{\`e}re}, {Gillard}, \&
  {Jean}}]{Ferriere:2007}
{Ferri{\`e}re}, K., {Gillard}, W., \& {Jean}, P. 2007, \aap, 467, 611

\bibitem[{{Gaensler} {et~al.}(2008){Gaensler}, {Madsen}, {Chatterjee}, \&
  {Mao}}]{Gaensler:2008}
{Gaensler}, B.~M., {Madsen}, G.~J., {Chatterjee}, S., \& {Mao}, S.~A. 2008,
  \pasa, 25, 184

\bibitem[{{Ginzburg}(1979)}]{Ginzburg:1979}
{Ginzburg}, V.~L. 1979, Oxford Pergamon Press International Series on Natural
  Philosophy, 99

\bibitem[{{Gordon} \& {Burton}(1976)}]{Gordon:1976}
{Gordon}, M.~A. \& {Burton}, W.~B. 1976, \apj, 208, 346

\bibitem[{{Guessoum} {et~al.}(2005){Guessoum}, {Jean}, \&
  {Gillard}}]{Guessoum:2005}
{Guessoum}, N., {Jean}, P., \& {Gillard}, W. 2005, \aap, 436, 171

\bibitem[{{Higdon} {et~al.}(2009){Higdon}, {Lingenfelter}, \&
  {Rothschild}}]{Higdon:2009}
{Higdon}, J.~C., {Lingenfelter}, R.~E., \& {Rothschild}, R.~E. 2009, \apj, 698,
  350

\bibitem[{{Jean} {et~al.}(2009){Jean}, {Gillard}, {Marcowith}, \&
  {Ferri{\`e}re}}]{Jean:2009}
{Jean}, P., {Gillard}, W., {Marcowith}, A., \& {Ferri{\`e}re}, K. 2009, \aap,
  508, 1099

\bibitem[{{Jean} {et~al.}(2006){Jean}, {Kn{\"o}dlseder}, {Gillard}, {Guessoum},
  {Ferri{\`e}re}, {Marcowith}, {Lonjou}, \& {Roques}}]{Jean:2006}
{Jean}, P., {Kn{\"o}dlseder}, J., {Gillard}, W., {et~al.} 2006, \aap, 445, 579

\bibitem[{{Kn{\"o}dlseder} {et~al.}(1999){Kn{\"o}dlseder}, {Bennett},
  {Bloemen}, {Diehl}, {Hermsen}, {Oberlack}, {Ryan}, {Sch{\"o}nfelder}, \& {von
  Ballmoos}}]{Knodlseder:1999}
{Kn{\"o}dlseder}, J., {Bennett}, K., {Bloemen}, H., {et~al.} 1999, \aap, 344,
  68

\bibitem[{{Lair} {et~al.}(2006){Lair}, {Leising}, {Milne}, \&
  {Williams}}]{Lair:2006}
{Lair}, J.~C., {Leising}, M.~D., {Milne}, P.~A., \& {Williams}, G.~G. 2006,
  \aj, 132, 2024

\bibitem[{{Leising} \& {Share}(1990)}]{Leising:1990}
{Leising}, M.~D. \& {Share}, G.~H. 1990, \apj, 357, 638

\bibitem[{{Lithwick} \& {Goldreich}(2001)}]{Lithwick:2001}
{Lithwick}, Y. \& {Goldreich}, P. 2001, \apj, 562, 279

\bibitem[{{Martin} {et~al.}(2009){Martin}, {Kn{\"o}dlseder}, {Diehl}, \&
  {Meynet}}]{Martin:2009a}
{Martin}, P., {Kn{\"o}dlseder}, J., {Diehl}, R., \& {Meynet}, G. 2009, \aap,
  506, 703

\bibitem[{{Martin} {et~al.}(2010){Martin}, {Vink}, {Jiraskova}, {Jean}, \&
  {Diehl}}]{Martin:2010}
{Martin}, P., {Vink}, J., {Jiraskova}, S., {Jean}, P., \& {Diehl}, R. 2010,
  \aap, 519, A100+

\bibitem[{{Milne} {et~al.}(1999){Milne}, {The}, \& {Leising}}]{Milne:1999}
{Milne}, P.~A., {The}, L.-S., \& {Leising}, M.~D. 1999, \apjs, 124, 503

\bibitem[{{Motizuki} \& {Kumagai}(2004)}]{Motizuki:2004}
{Motizuki}, Y. \& {Kumagai}, S. 2004, in American Institute of Physics
  Conference Series, Vol. 704, Tours Symposium on Nuclear Physics V, ed.
  M.~{Arnould}, M.~{Lewitowicz}, G.~{M{\"u}nzenberg}, H.~{Akimune}, M.~{Ohta},
  H.~{Utsunomiya}, T.~{Wada}, \& T.~{Yamagata}, 369--374

\bibitem[{{Ore} \& {Powell}(1949)}]{Ore:1949}
{Ore}, A. \& {Powell}, J.~L. 1949, Physical Review, 75, 1696

\bibitem[{{Pages} {et~al.}(1972){Pages}, {Bertel}, {Joffre}, \&
  {Sklavenitis}}]{Pages:1972}
{Pages}, L., {Bertel}, E., {Joffre}, H., \& {Sklavenitis}, L. 1972, Atomic
  Data, 4, 1

\bibitem[{{Pl{\"u}schke} {et~al.}(2001){Pl{\"u}schke}, {Diehl},
  {Sch{\"o}nfelder}, {Bloemen}, {Hermsen}, {Bennett}, {Winkler}, {McConnell},
  {Ryan}, {Oberlack}, \& {Kn{\"o}dlseder}}]{Pluschke:2001}
{Pl{\"u}schke}, S., {Diehl}, R., {Sch{\"o}nfelder}, V., {et~al.} 2001, in ESA
  Special Publication, Vol. 459, Exploring the Gamma-Ray Universe, ed.
  A.~{Gimenez}, V.~{Reglero}, \& C.~{Winkler}, 55--58

\bibitem[{{Porter} {et~al.}(2008){Porter}, {Moskalenko}, {Strong}, {Orlando},
  \& {Bouchet}}]{Porter:2008}
{Porter}, T.~A., {Moskalenko}, I.~V., {Strong}, A.~W., {Orlando}, E., \&
  {Bouchet}, L. 2008, \apj, 682, 400

\bibitem[{{Prantzos}(2004)}]{Prantzos:2004}
{Prantzos}, N. 2004, in ESA Special Publication, Vol. 552, 5th INTEGRAL
  Workshop on the INTEGRAL Universe, ed. V.~{Schoenfelder}, G.~{Lichti}, \&
  C.~{Winkler}, 15--+

\bibitem[{{Prantzos}(2006)}]{Prantzos:2006}
{Prantzos}, N. 2006, \aap, 449, 869

\bibitem[{{Prantzos} {et~al.}(2011){Prantzos}, {Boehm}, {Bykov}, {Diehl},
  {Ferri{\`e}re}, {Guessoum}, {Jean}, {Knoedlseder}, {Marcowith}, {Moskalenko},
  {Strong}, \& {Weidenspointner}}]{Prantzos:2011}
{Prantzos}, N., {Boehm}, C., {Bykov}, A.~M., {et~al.} 2011, Reviews of Modern
  Physics, 83, 1001

\bibitem[{{Prantzos} \& {Diehl}(1996)}]{Prantzos:1996}
{Prantzos}, N. \& {Diehl}, R. 1996, \physrep, 267, 1

\bibitem[{{Ptuskin} {et~al.}(2006){Ptuskin}, {Moskalenko}, {Jones}, {Strong},
  \& {Zirakashvili}}]{Ptuskin:2006}
{Ptuskin}, V.~S., {Moskalenko}, I.~V., {Jones}, F.~C., {Strong}, A.~W., \&
  {Zirakashvili}, V.~N. 2006, \apj, 642, 902

\bibitem[{{Ragot}(2006)}]{Ragot:2006}
{Ragot}, B.~R. 2006, \apj, 642, 1163

\bibitem[{{Renaud} {et~al.}(2006){Renaud}, {Vink}, {Decourchelle}, {Lebrun},
  {Hartog}, {Terrier}, {Couvreur}, {Kn{\"o}dlseder}, {Martin}, {Prantzos},
  {Bykov}, \& {Bloemen}}]{Renaud:2006}
{Renaud}, M., {Vink}, J., {Decourchelle}, A., {et~al.} 2006, \apjl, 647, L41

\bibitem[{{Robin} {et~al.}(2003){Robin}, {Reyl{\'e}}, {Derri{\`e}re}, \&
  {Picaud}}]{Robin:2003}
{Robin}, A.~C., {Reyl{\'e}}, C., {Derri{\`e}re}, S., \& {Picaud}, S. 2003,
  \aap, 409, 523

\bibitem[{{Sizun} {et~al.}(2006){Sizun}, {Cass{\'e}}, \&
  {Schanne}}]{Sizun:2006}
{Sizun}, P., {Cass{\'e}}, M., \& {Schanne}, S. 2006, \prd, 74, 063514

\bibitem[{{Strong} {et~al.}(2007){Strong}, {Moskalenko}, \&
  {Ptuskin}}]{Strong:2007}
{Strong}, A.~W., {Moskalenko}, I.~V., \& {Ptuskin}, V.~S. 2007, Annual Review
  of Nuclear and Particle Science, 57, 285

\bibitem[{{Strong} {et~al.}(2011){Strong}, {Orlando}, \& {Jaffe}}]{Strong:2011}
{Strong}, A.~W., {Orlando}, E., \& {Jaffe}, T.~R. 2011, \aap, 534, A54

\bibitem[{{Strong} {et~al.}(2010){Strong}, {Porter}, {Digel},
  {J{\'o}hannesson}, {Martin}, {Moskalenko}, {Murphy}, \&
  {Orlando}}]{Strong:2010}
{Strong}, A.~W., {Porter}, T.~A., {Digel}, S.~W., {et~al.} 2010, \apjl, 722,
  L58

\bibitem[{{Sullivan} {et~al.}(2006){Sullivan}, {Le Borgne}, {Pritchet},
  {Hodsman}, {Neill}, {Howell}, {Carlberg}, {Astier}, {Aubourg}, {Balam},
  {Basa}, {Conley}, {Fabbro}, {Fouchez}, {Guy}, {Hook}, {Pain},
  {Palanque-Delabrouille}, {Perrett}, {Regnault}, {Rich}, {Taillet}, {Baumont},
  {Bronder}, {Ellis}, {Filiol}, {Lusset}, {Perlmutter}, {Ripoche}, \&
  {Tao}}]{Sullivan:2006}
{Sullivan}, M., {Le Borgne}, D., {Pritchet}, C.~J., {et~al.} 2006, \apj, 648,
  868

\bibitem[{{Tammann} {et~al.}(1994){Tammann}, {Loeffler}, \&
  {Schroeder}}]{Tammann:1994}
{Tammann}, G.~A., {Loeffler}, W., \& {Schroeder}, A. 1994, \apjs, 92, 487

\bibitem[{{The} {et~al.}(2006){The}, {Clayton}, {Diehl}, {Hartmann}, {Iyudin},
  {Leising}, {Meyer}, {Motizuki}, \& {Sch{\"o}nfelder}}]{The:2006}
{The}, L.-S., {Clayton}, D.~D., {Diehl}, R., {et~al.} 2006, \aap, 450, 1037

\bibitem[{{Wang} {et~al.}(2009){Wang}, {Lang}, {Diehl}, {Halloin}, {Jean},
  {Kn{\"o}dlseder}, {Kretschmer}, {Martin}, {Roques}, {Strong}, {Winkler}, \&
  {Zhang}}]{Wang:2009}
{Wang}, W., {Lang}, M.~G., {Diehl}, R., {et~al.} 2009, \aap, 496, 713

\bibitem[{{Weidenspointner} {et~al.}(2008{\natexlab{a}}){Weidenspointner},
  {Skinner}, {Jean}, {Kn{\"o}dlseder}, {von Ballmoos}, {Bignami}, {Diehl},
  {Strong}, {Cordier}, {Schanne}, \& {Winkler}}]{Weidenspointner:2008}
{Weidenspointner}, G., {Skinner}, G., {Jean}, P., {et~al.} 2008{\natexlab{a}},
  \nat, 451, 159

\bibitem[{{Weidenspointner} {et~al.}(2008{\natexlab{b}}){Weidenspointner},
  {Skinner}, {Jean}, {Kn{\"o}dlseder}, {von Ballmoos}, {Diehl}, {Strong},
  {Cordier}, {Schanne}, \& {Winkler}}]{Weidenspointner:2008a}
{Weidenspointner}, G., {Skinner}, G.~K., {Jean}, P., {et~al.}
  2008{\natexlab{b}}, New Astronomy Review, 52, 454

\bibitem[{{Wouterloot} {et~al.}(1990){Wouterloot}, {Brand}, {Burton}, \&
  {Kwee}}]{Wouterloot:1990}
{Wouterloot}, J.~G.~A., {Brand}, J., {Burton}, W.~B., \& {Kwee}, K.~K. 1990,
  \aap, 230, 21

\bibitem[{{Yan} \& {Lazarian}(2004)}]{Yan:2004}
{Yan}, H. \& {Lazarian}, A. 2004, \apj, 614, 757

\bibitem[{{Zirakashvili} \& {Aharonian}(2011)}]{Zirakashvili:2011}
{Zirakashvili}, V.~N. \& {Aharonian}, F.~A. 2011, \prd, 84, 083010

\end{thebibliography}

\end{document}